\documentclass[aps,prb,reprint,floatfix]{revtex4-2}
\usepackage{graphicx}
\usepackage{tikz}
\usepackage{hyperref}
\usepackage{physics}
\usepackage{amsfonts,amsmath}

\begin{document}

\title{Simulation of three-dimensional quantum systems with projected entangled-pair states}

\author{Patrick C.G. Vlaar}
\email[]{p.c.g.vlaar@uva.nl}
\affiliation{Institute for Theoretical Physics and Delta Institute for Theoretical Physics,
             University of Amsterdam, Science Park 904, 1098 XH Amsterdam, The Netherlands}

\author{Philippe Corboz}
\affiliation{Institute for Theoretical Physics and Delta Institute for Theoretical Physics,
             University of Amsterdam, Science Park 904, 1098 XH Amsterdam, The Netherlands}

\date{\today}

\begin{abstract}
Tensor network algorithms have proven to be very powerful tools for studying one- and two-dimensional quantum many-body systems. However, their application to three-dimensional (3D) quantum systems has so far been limited, mostly because the efficient contraction of a 3D tensor network is very challenging. In this paper we develop and benchmark two contraction approaches for infinite projected entangled-pair states (iPEPS) in 3D. The first approach is based on a contraction of a finite cluster of tensors including an effective environment to approximate the full 3D network. The second approach performs a full contraction of the network by first iteratively contracting layers of the network with a boundary iPEPS, followed by a contraction of the resulting quasi-2D network using the corner transfer matrix renormalization group. Benchmark data for the Heisenberg and Bose-Hubbard models on the cubic lattice show that the algorithms provide competitive results compared to other approaches, making iPEPS a promising tool to study challenging open problems in 3D.
\end{abstract}

\maketitle

\section{Introduction}

The study of strongly correlated quantum systems has proven to be one of the most challenging endeavors in modern physics. Understanding the emergent phenomena in these systems often requires a synergy between innovative methods from theory, numerics, and experiments.
One important family of numerical methods that has seen rapid developments in recent decades are tensor network algorithms. These techniques have in common that the wavefunction is approximated by a variational ansatz formed by a product of tensors, where the accuracy is systematically controlled by the bond dimension of the tensors. The best known example is the matrix product state (MPS)~\cite{Fannes1992,Ostlund1995}, the one-dimensional (1D) ansatz formed by a product of rank-3 tensors, which is the underlying variational state of the powerful density matrix renormalization group (DMRG)~\cite{White1992,Schollwock2011} method. 
  
The projected entangled-pair state (PEPS)~\cite{Verstraete2004,Murg2007} (also known as tensor product state~\cite{Nishio2004}) was introduced as a higher-dimensional generalization of MPS, enabling the representation of ground states of large 2D systems,  or even infinite 2D systems, called infinite PEPS (iPEPS)~\cite{Jordan2008}. Thanks to significant progress on the algorithmic side over the past years, (i)PEPS has become a powerful approach for two-dimensional (2D) strongly correlated systems, especially for 2D fermionic and frustrated systems which are notoriously hard to simulate with quantum Monte Carlo (QMC), see e.g. Refs.~\cite{Corboz2014(2),Corboz2014,Liao2017,Niesen2017,Zheng2017,haghshenas18,Ponsioen2019,chung19,kshetrimayum20,shi19,lee20,chen20,liu20,hasik20}. Besides the computation of ground states, for which (i)PEPS was originally developed, significant progress has also been achieved in other applications, including the study of thermodynamic properties~\cite{Li2011,Ran2012,Czarnik2012,Ran2013,Czarnik2015(2),Czarnik2015,Peng2017,Dai2017,Czarnik2017,Ran2018,Czarnik2018,Chen2018,kshetrimayum19,Czarnik2019,Ran2019,wietek19,czarnik19b,czarnik19c,jimenez20,czarnik20}, excited states~\cite{Vanderstraeten2019,Ponsioen2020}, real-time evolution~\cite{Czarnik2019,Hubig2019,Hubig2020,Kshetrimayum2020} and open systems~\cite{Kshetrimayum2017,Czarnik2019}.

The successes of tensor networks for both 1D and 2D quantum systems raise the question of whether these methods can also be applied to three-dimensional (3D) quantum systems. It is expected that ground states in 3D  typically require a smaller bond dimension than their lower-dimensional counterparts, because the entanglement of a site gets shared with more neighbors, such that 3D states typically lie closer to a product state. On the other hand, the presence of additional legs on the tensors  implies a higher computational cost of the algorithms,  forming one of the main obstacles in their development.

The main challenge of higher dimensional tensor networks is that they cannot be  contracted exactly, but only approximately, in contrast to the MPS. Several 2D contraction approaches have been developed, which can roughly be divided into three categories: the corner transfer matrix renormalization group (CTMRG)~\cite{Nishino1996,Orus2009,Corboz2014,Fishman2018}, the (higher-order) tensor renormalization group (TRG)~\cite{Levin2007,Xie2009,Xie2012} and the related tensor network renormalization (TNR)~\cite{Evenbly2015,Evenbly2017}, and boundary MPS algorithms~\cite{Vidal2004,Verstraete2004,Jordan2008,Zauner-Stauber2018,Fishman2018}. Several generalizations of these approaches to 3D have been proposed, including CTMRG in 3D~\cite{Nishino1998} and TRG-based methods, such as the higher-order TRG (HOTRG)~\cite{Xie2012} and other algorithms~\cite{Garcia-Saez2013,Teng2017}. 3D generalizations of the third category, based on a boundary iPEPS, have been introduced for 3D classical systems~\cite{Nishino2000,Nishino2001,Gendiar2002,Gendiar2003,Gendiar2005,Vanderstraeten2018}, or also in the context of imaginary time evolution algorithms of 2D quantum states at zero~\cite{Jiang2008,Jordan2008,Phien2015} and finite temperature~\cite{Li2011,Ran2012,Czarnik2012,Ran2013,Czarnik2015(2),Czarnik2015,Czarnik2018,Czarnik2019}, which effectively corresponds to a contraction of an anisotropic 3D network. 
Outside of these broad categories other algorithms were  proposed for 3D networks, including embedding a small bulk part in an entanglement bath~\cite{Ran2017,Ran2019}, graph-based PEPS~\cite{Jahromi2019,Jahromi2020}, and isometric tensor networks~\cite{Tepaske2020}.

In this paper we extend the algorithmic toolbox by proposing two contraction techniques for the study of 3D quantum models. The first method is based on the exact contraction of only a finite number of tensors while using an effective environment to approximate the rest of the network. This approach, which we call cluster contraction in the following, provides a simple, approximate contraction, with the accuracy being controlled by the cluster size. It forms an extension of the approximate two-site cluster contraction used in previous works~\cite{Picot2015,Picot2016,Jahromi2019,Jahromi2020}, and a related idea was also used in the cluster update and evaluation procedures proposed in Refs.~\cite{Wang2011,Lubasch2014,Lubasch2014(2)} in 2D. 

In the second approach a full contraction of the network is performed by defining a 2D boundary iPEPS and by iteratively absorbing layers of the 3D network (which can be seen as infinite projected entangled-pair operator (iPEPO) layers) into the boundary iPEPS. The contraction is based on the simple update (SU) scheme~\cite{Jiang2008} which is commonly used in imaginary time evolution algorithms. After convergence of the boundary iPEPS, the 3D tensor network can be effectively represented by a quasi-2D network which is contracted using the CTMRG algorithm. The accuracy of this SU+CTMRG algorithm can be systematically controlled by the bond dimensions of the boundary iPEPS and CTMRG environment tensors. 

The paper is organized as follows: In Sec.~\ref{sec:intro_ipeps} a short introduction to the iPEPS ansatz and the  optimization algorithm based on the 3D SU imaginary time evolution algorithm is given. Then, the cluster contractions and the SU+CTMRG method are introduced in Sec.~\ref{sec:cluster_contractions} and Sec.~\ref{sec:SU+CTMRG}, respectively.  In Sec.~\ref{sec:results} benchmark results for the Heisenberg  and the Bose-Hubbard model on the cubic lattice are provided, with a comparison to previous studies based on QMC and other approaches. Finally, we present our conclusions and outlook in Sec.~\ref{sec:discussion}.

\section{Introduction to iPEPS} \label{sec:intro_ipeps}

\subsection{iPEPS ansatz} \label{sec:ipeps_ansatz}

A PEPS is a variational ansatz which approximates the wavefunction as a trace over a product of tensors. It forms a natural generalization of the 1D MPS to higher dimensional systems~\cite{Verstraete2004}. A general way to define the ansatz is
\begin{equation}
	\ket{\psi} = \sum_{s_1\ldots s_N=1}^d \Tr(T^{\vec{r}_1}_{s_1} \ldots T^{\vec{r}_N}_{s_N}) \ket{s_1 \ldots s_N} \label{eq:general_peps}
\end{equation}
where $\vec{r}_i$ indicates the position of the tensor $T_{s_i}^{\vec{r}_i}$ in the ansatz and $s_i$ represents the index of the local Hilbert space of a site. The tensors are connected with each other, typically according to the underlying lattice. 
By defining a supercell of tensors and repeating this cell infinitely many times on the lattice we obtain an infinite PEPS (iPEPS), which  represents a wavefunction directly in the thermodynamic limit~\cite{Jordan2008}. In this work we limit ourselves to the study of cubic lattices with two independent tensors on the two sublattices, where each tensor has six $D$-dimensional auxiliary indices connecting each tensor with its nearest neighbors and one $d$-dimensional physical index carrying the local Hilbert space. The parameter $D$ is called the bond dimension and controls the accuracy of the ansatz. 

A ground state simulation using iPEPS consists of two stages. First, the ansatz is optimized such that it forms an accurate representation of the ground state of some given Hamiltonian. The optimization of the iPEPS is commonly done either by an energy minimization~\cite{Verstraete2004,Corboz2016,Vanderstraeten2016,Liao2019} or by an imaginary time evolution~\cite{Jiang2008,Jordan2008,Phien2015}. Once the iPEPS has been optimized, properties of the state, such as expectation values of observables, can be computed. In general both stages involve a contraction of the 3D tensor network. However, for the optimization we will make use of the SU imaginary time evolution algorithm~\cite{Jiang2008} which is a local, approximate optimization scheme that does not require a full contraction. This algorithm is discussed in the following for  the 3D case, before turning our focus on the  contraction methods in Secs.~\ref{sec:cluster_contractions} and~\ref{sec:SU+CTMRG}.

\subsection{Simple update imaginary time evolution} \label{sec:simple_update}

The main idea of an imaginary time evolution algorithm is to project an initial state $\ket{\phi}$ onto the ground state $\ket{\psi_0}$ by acting with the imaginary time evolution operator $e^{-\beta \hat{H}}$ in the infinite $\beta$ limit, 
\begin{equation}
	e^{-\beta \hat{H}} \ket{\phi} \overset{\beta\rightarrow\infty}{\rightarrow} \ket{\psi_0}
\end{equation}
where $\hat H$ is the Hamiltonian. In practice, the imaginary time evolution operator  is split up into smaller two-body gates using a Trotter-Suzuki decomposition. The first-order Trotter-Suzuki decomposition is given by
\begin{equation}
	e^{-\beta\sum_i \hat{H}_i} = \left(e^{-\tau\sum_i \hat{H}_i} \right)^M = \prod_{j=1}^{M} \prod_i e^{-\tau\hat{H}_i} + \mathcal{O}(\tau),
\end{equation}
where the Hamiltonian is rewritten as a sum over local terms $\hat{H} = \sum_i \hat{H}_i$ and $\tau=\beta/M$ is the time step. The error can be reduced further to $\mathcal{O}(\tau^2)$ by reverting the sequence of applied gates $e^{-\tau\hat{H}_i}$ in every other time step, corresponding to a second-order Trotter-Suzuki decomposition which will be used here. 

Upon absorbing a two-body gate, the bond dimension of the ansatz grows from $D$ to $d^2D$, which must be truncated to avoid an exponential growth of the bond dimension. There exist different truncation schemes to achieve this. In the full~\cite{Jordan2008} (or fast-full~\cite{Phien2015}) update approach the entire wave function is taken into account to truncate a bond index, which is optimal but computationally expensive, since it requires a contraction of the network at each iteration. In contrast, in the SU approach the truncation is done by a local singular value decomposition (SVD) which is not as accurate as the full update, but computationally substantially cheaper, and thus we focus on this approach in the following. 

\begin{figure}
	\includegraphics[width=0.9652\linewidth]{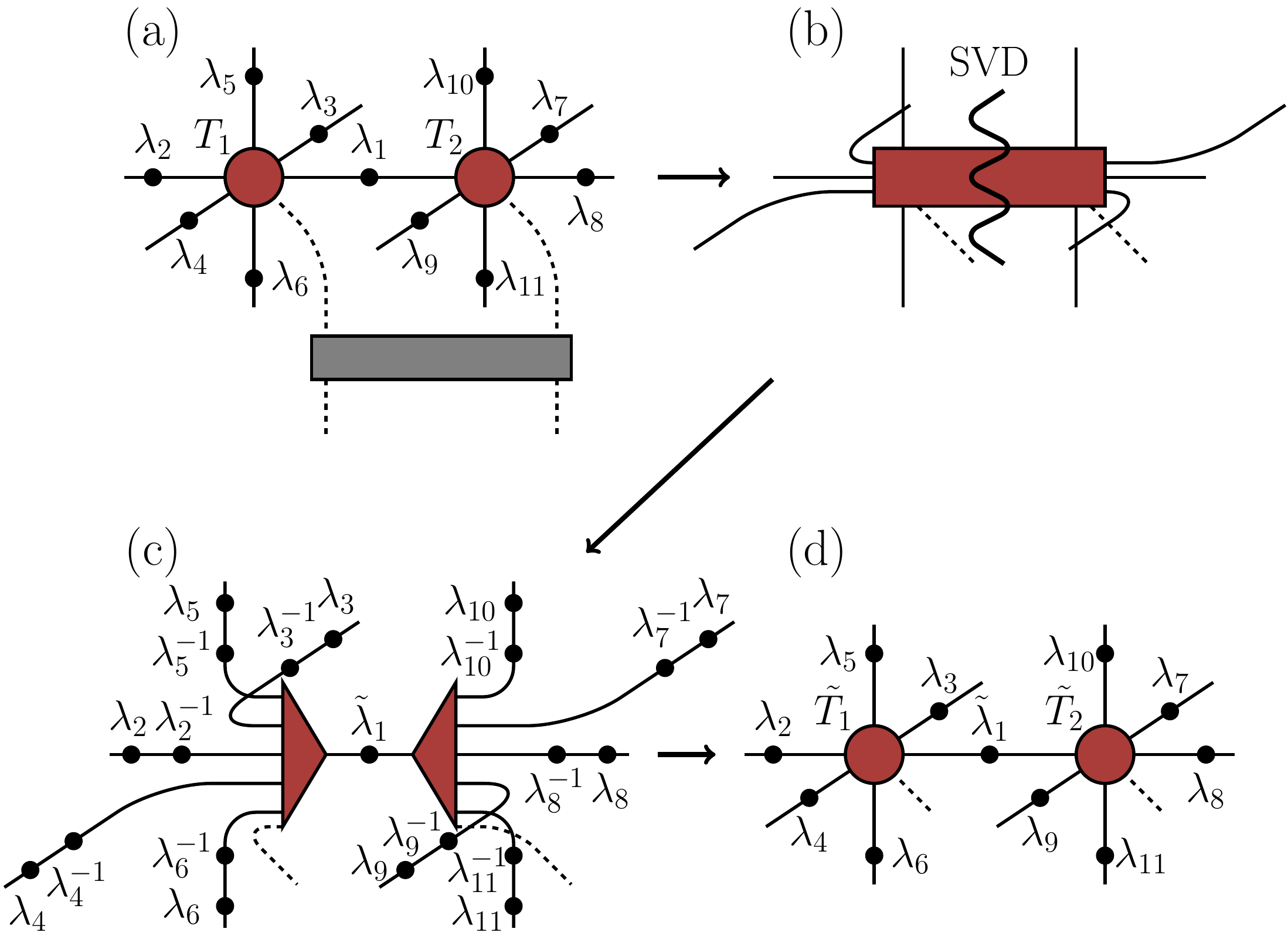}
	\caption{One elementary step in the SU imaginary time evolution algorithm. In (a) a Trotter gate is multiplied onto two neighboring tensors, including all adjacent singular value matrices $\lambda_i$ which are used as an approximate environment of the bond. In (b) the resulting tensor is split, and the connecting bond is truncated using an SVD. In (c) the singular value matrices are reintroduced on the outward pointing auxiliary bonds by inserting identities $\lambda_i \lambda_i^{-1}$. In (d) the $\lambda_i^{-1}$ matrices are absorbed, giving the updated tensors.}
	\label{fig:su_step}
\end{figure}

The scheme is presented in Fig.~\ref{fig:su_step} for the 3D cubic network, which is a straightforward generalization of the 2D case~\cite{Jiang2008}. The ansatz defined in Eq.~\eqref{eq:general_peps} is altered by adding diagonal, positive-valued matrices $\lambda_i$ on the auxiliary bonds which are obtained from the SVD. In the case of an MPS, absorbing the singular value matrices adjacent to two tensors connected by a bond brings the MPS into a canonical form with respect to that bond, such that the SVD provides an optimal truncation~\cite{Vidal2007}. In 2D or 3D there is no canonical form because of the loops in the tensor network ansatz. Still, absorbing the singular value matrices adjacent to a bond brings that bond close to a canonical form (more specifically a quasicanonical~\cite{Kalis2012,Phien2015(2)} or superorthogonal~\cite{Ran2012} form), and the local SVD often yields a truncation with a good accuracy. Note that the original representation of Eq.~\eqref{eq:general_peps} can be recovered by absorbing $\sqrt{\lambda_i}$ on all sides of a tensor.
The computational cost of the SU can be reduced further by splitting off the physical bond from the tensors using a QR decomposition before absorbing the Trotter gate~\cite{Corboz2010}. 

While the SU update is used here to truncate a bond index within the imaginary time evolution algorithm, we will make use of the same idea for the contraction of the 3D tensor network in Sec.~\ref{sec:SU+CTMRG}. Furthermore, using the singular value matrices as effective environments also plays a key role in the cluster contraction, which we discuss in the following.

\section{Cluster contraction} \label{sec:cluster_contractions}

\begin{figure}
	\includegraphics[width=0.4087\linewidth]{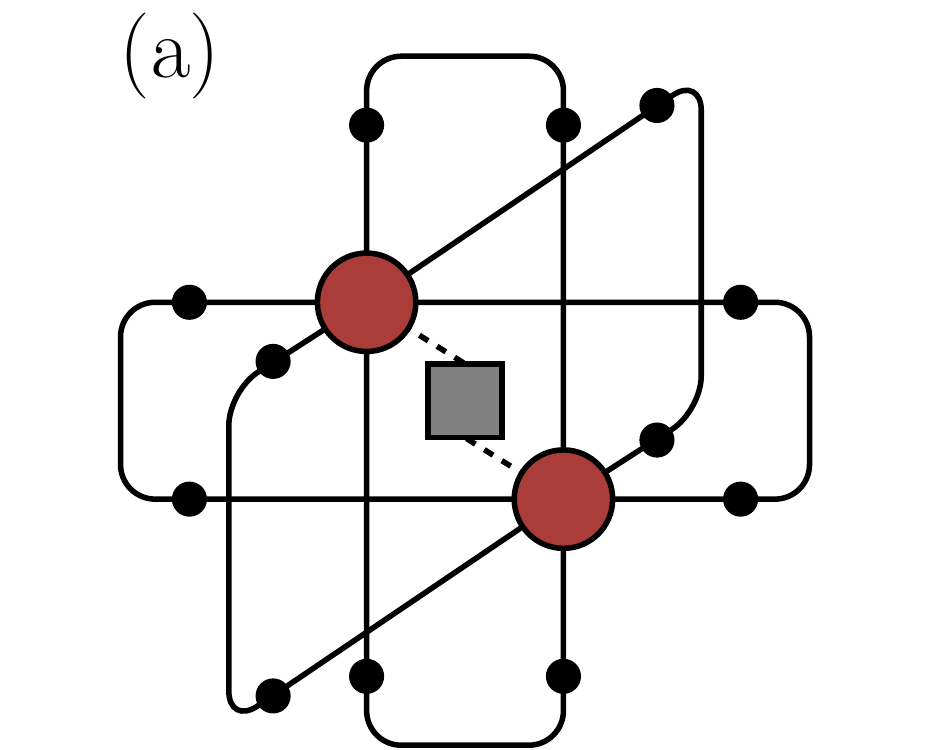}
	\includegraphics[width=0.5870\linewidth]{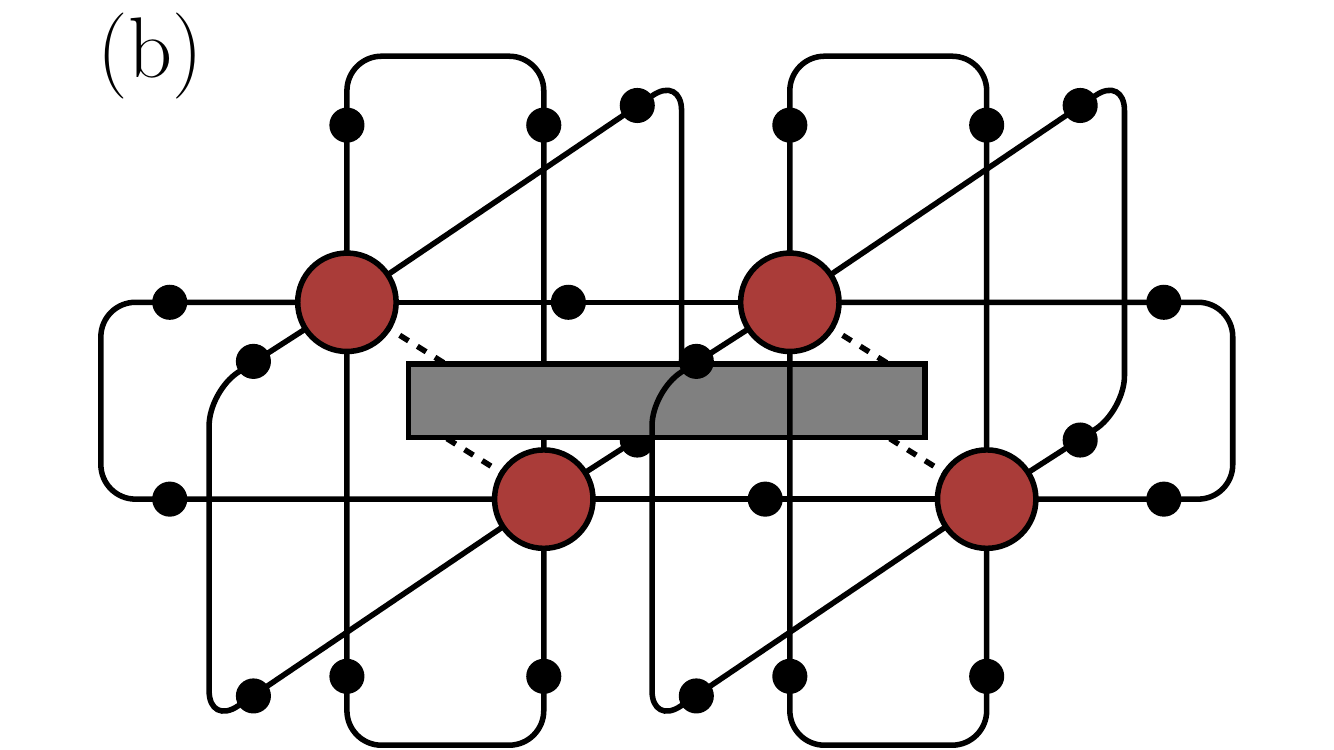}
	\caption{Diagrams of the smallest clusters used in the cluster contraction. In (a) the $1\times1\times1$ cluster is displayed which is used to evaluate one-site operators and in (b) the $1\times1\times2$ cluster is shown which is used for two-site operators. The black circles represent singular value matrices, which provide an effective environment, approximating the rest of the tensor network surrounding the cluster.}
	\label{fig:smallest_clusters}
\end{figure}

We will now shift the attention to computing expectation values of the optimized iPEPS. The first contraction method that is introduced is the cluster contraction. In this approach, instead of performing a full contraction of the network, only a small cluster of the network is contracted exactly, while the rest of the network is taken into account only in an approximate way by absorbing the singular value matrices on the outer legs of the cluster, in a similar spirit as done in the SU imaginary time evolution algorithm discussed in the previous section. The smallest clusters are the $1\times1\times1$ and $1\times1\times2$ clusters depicted in Fig.~\ref{fig:smallest_clusters} which can be used to evaluate one- and two-site operators respectively. These contractions have a relatively low computational cost of $\mathcal{O}(D^7)$ and they were used as an approximate contraction method before~\cite{Picot2015,Picot2016,Jahromi2019,Jahromi2020}. We note that these contractions are exact on a Bethe lattice (containing no loops), and have been used frequently in this context~\cite{Shi2006,Nagaj2008,Nagy2012,Li2012,Depenbrock2013}.

\begin{figure}
	\includegraphics[width=0.3696\linewidth]{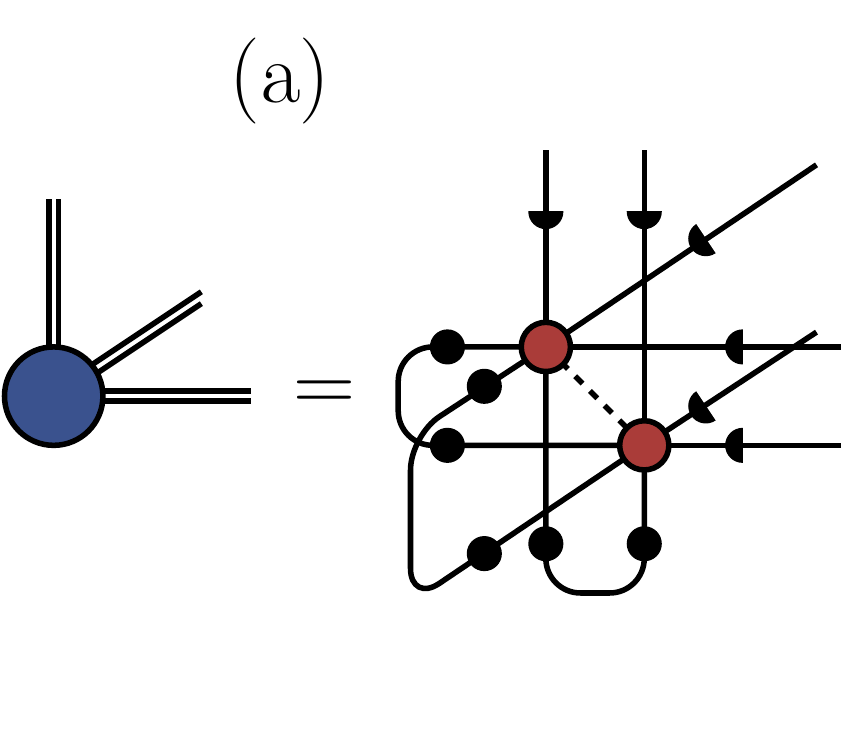}\hspace{.05\linewidth}
	\includegraphics[width=0.5652\linewidth]{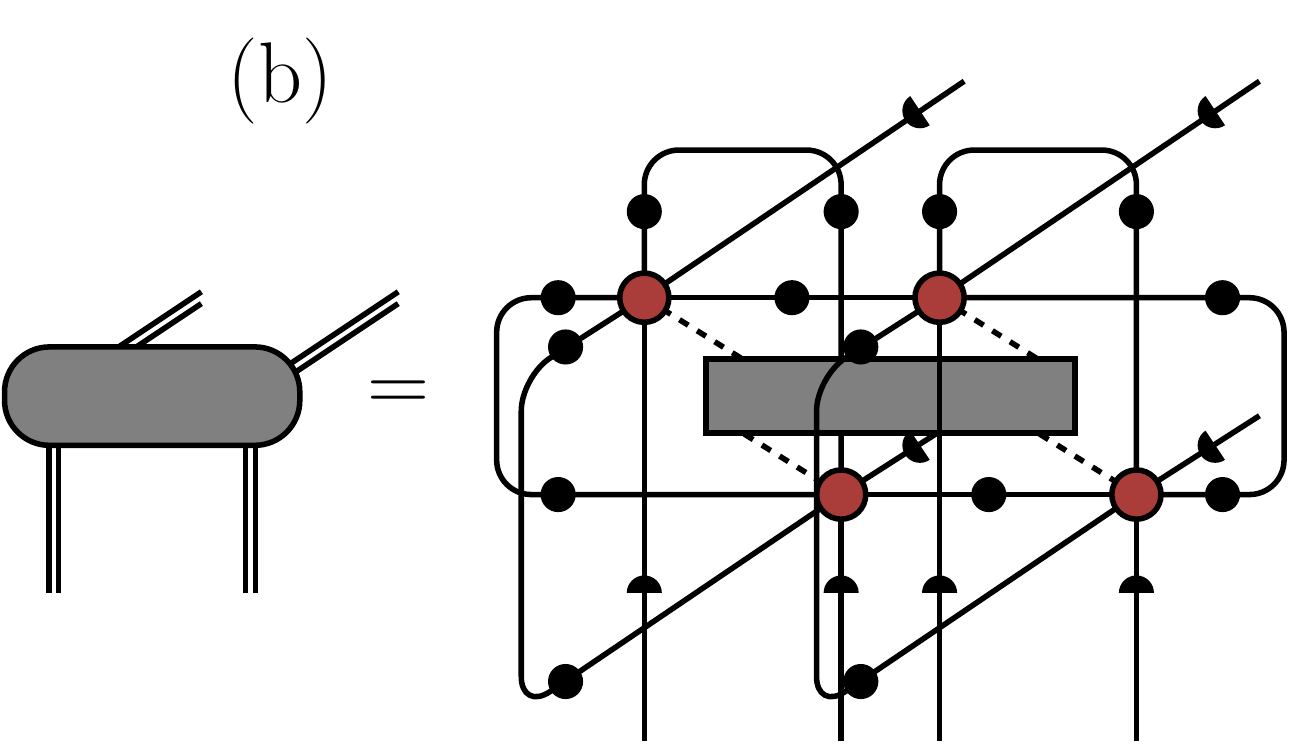}\vspace{.01\textheight}
	\includegraphics[width=0.3696\linewidth]{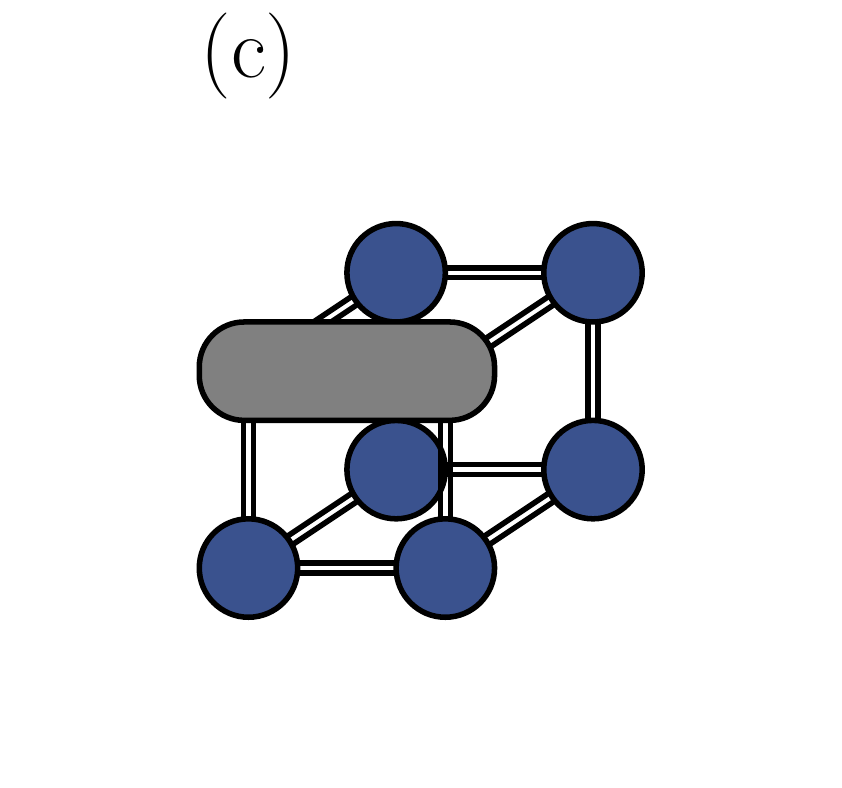}\hspace{.05\linewidth}
	\includegraphics[width=0.5652\linewidth]{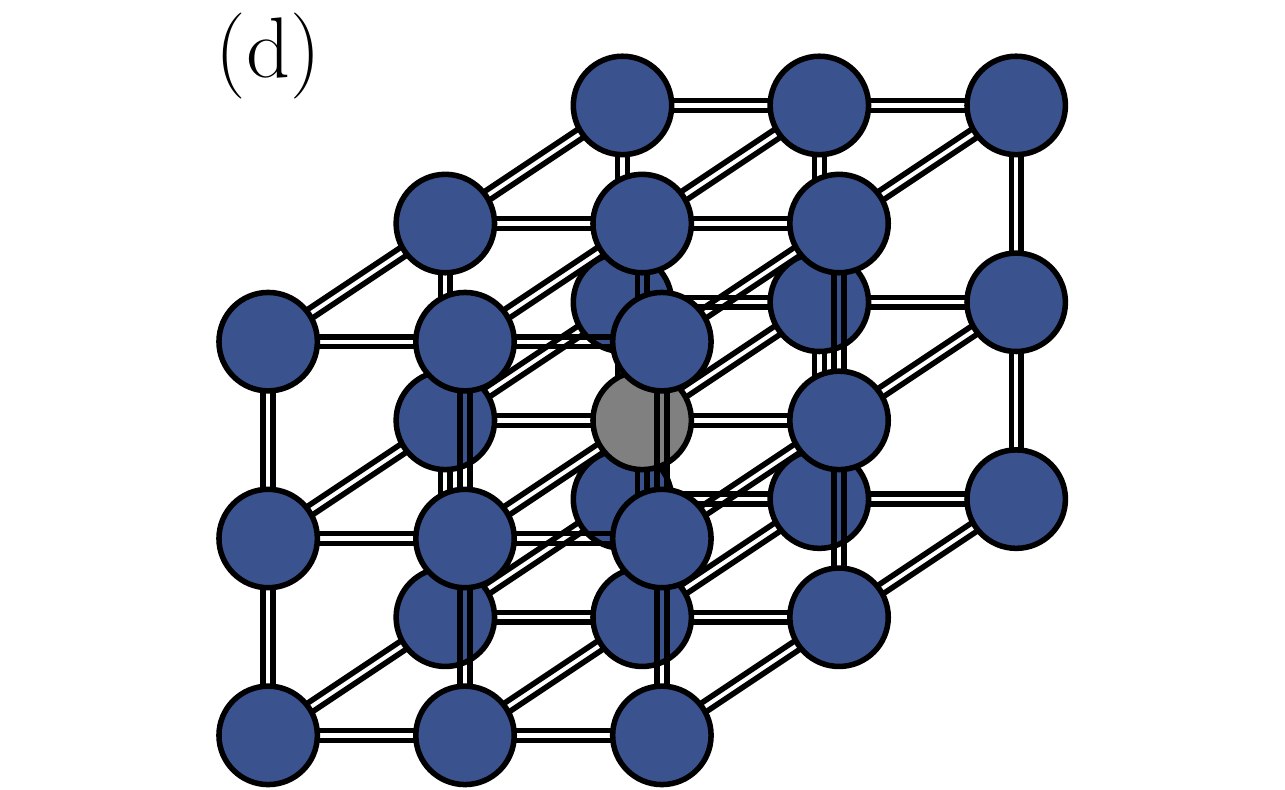}\vspace{.01\textheight}
	\includegraphics[width=0.5478\linewidth]{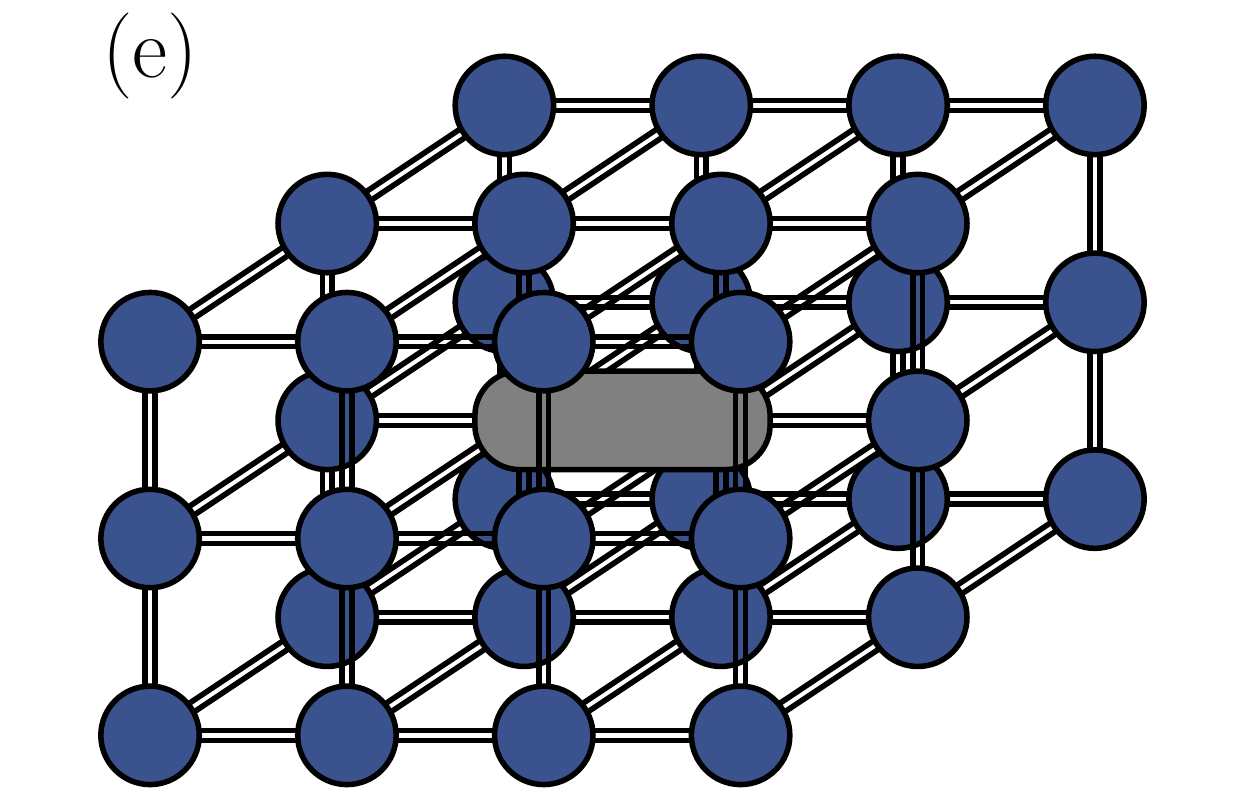}
	\caption{Cluster contractions for larger clusters. In (a) and (b) a more compact notation is introduced for graphical clarity. A full black circle represents $\lambda_i$, while a half circle represents $\sqrt{\lambda_i}$. Note that keeping the tensors separated typically gives a lower computational cost. In (c) the $2\times2\times2$ cluster is displayed. In practice separate diagrams, with the operator inserted on the different bonds of this cluster, are computed and averaged over. Diagrams (d) and (e) show the $3\times3\times3$ and $3\times3\times4$ clusters, respectively.}
	\label{fig:larger_clusters}
\end{figure}

While these small clusters have the advantage of having a low computational cost, the involved contraction error may be quite substantial because of their small sizes and because they entirely neglect loops in the network. Furthermore, without a systematic way of increasing the contraction accuracy it is hard to estimate the magnitude of the contraction error. For these reasons, we will extend the cluster contractions to larger clusters in this work. Adding an additional layer of tensors around the site(s) on which the operator is measured results in the $3\times3\times3$ and $3\times3\times4$ clusters, which are depicted in Fig.~\ref{fig:larger_clusters}(d-e). 
The computational cost of contracting these clusters has a high scaling of $\mathcal{O}(D^{29})$, therefore in practice it can typically only be used for $D=2$ (without approximations). In addition, we consider the $2\times2\times2$ cluster shown in Fig.~\ref{fig:larger_clusters}(c) with a contraction cost scaling as $\mathcal{O}(D^{12})$, which we find offers a good trade-off between accuracy and computational cost.

Cluster contractions are expected to provide reasonable results for states with short-ranged correlations. Their main advantage is that they are simple to implement and, for the smaller clusters, computationally relatively cheap to perform. They can therefore be used for a quick first analysis of a model and to identify parameter regions that could be interesting to simulate using more sophisticated methods, such as the SU+CTMRG contraction method we introduce in the following.

\section{SU+CTMRG contraction} \label{sec:SU+CTMRG}

\begin{figure}
	\includegraphics[width=\linewidth]{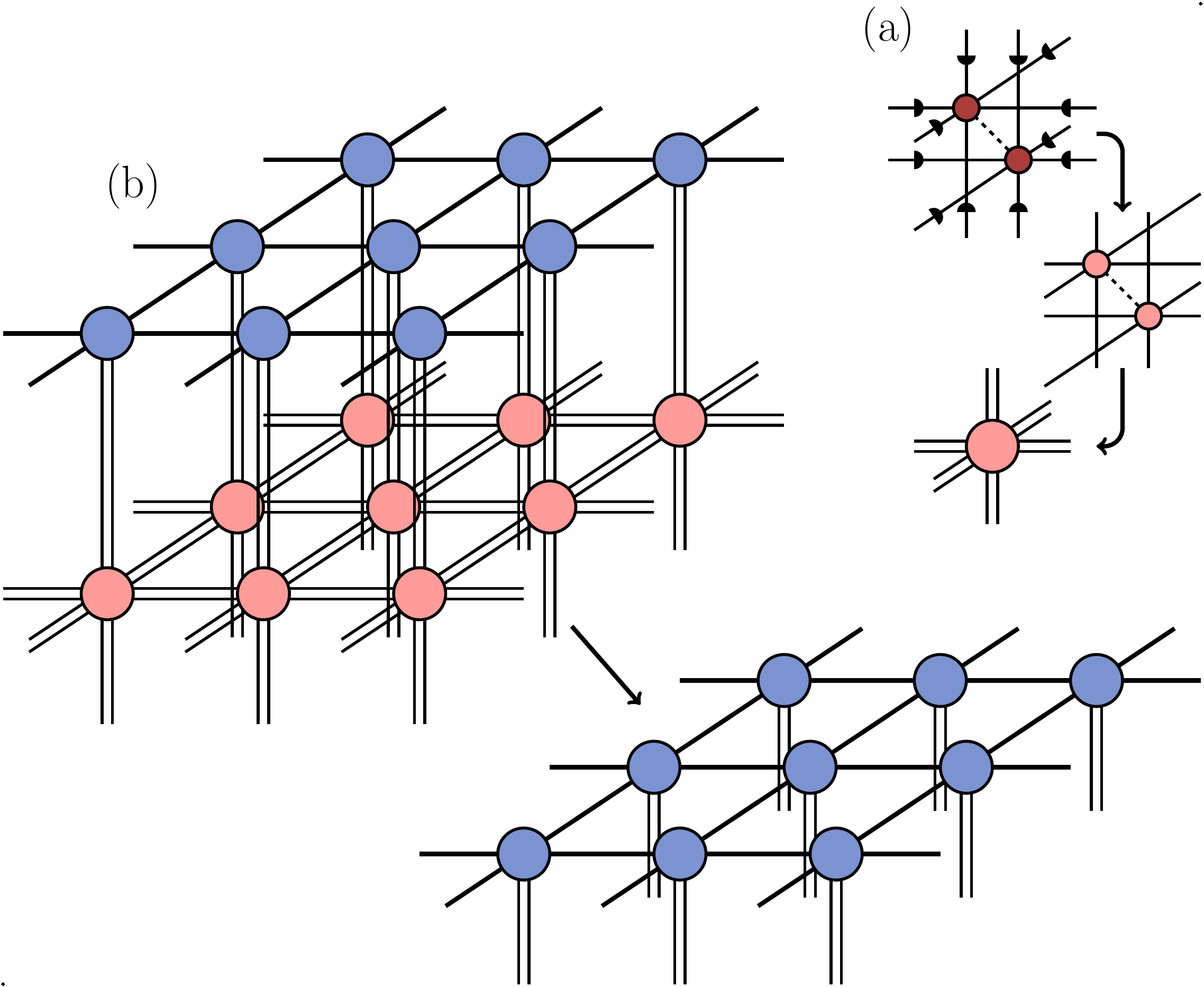}
	\caption{To contract the 3D tensor network a 2D boundary iPEPS is defined. For graphical clarity the bulk 3D iPEPS tensors are represented in the way shown in (a). The half circles represent $\sqrt{\lambda_i}$ which are contracted into the bulk tensors. In (b) a contraction of an iPEPO layer with the boundary iPEPS is displayed.}
	\label{fig:boundary_peps}
\end{figure}

In this section we introduce a method to perform a full contraction of the infinite 3D network with an accuracy that can be systematically controlled. The approach is based on a boundary iPEPS onto which layers of the 3D network (which can be seen as iPEPO's) are absorbed, see Fig.~\ref{fig:boundary_peps}, in a similar spirit as done in MPS-MPO (matrix product operator) based contractions of 2D tensor networks~\cite{Verstraete2004}. Methods based on a boundary iPEPS have been previously developed in the context of 3D classical models~\cite{Nishino2000,Nishino2001,Gendiar2002,Gendiar2003,Gendiar2005,Vanderstraeten2018}. While these approaches are typically based on a direct optimization of the boundary iPEPS, here we propose a computationally cheaper scheme, that is applicable also for general 3D tensor networks without any mirror or rotational symmetries. 

The method is based on iterative absorptions of iPEPO layers onto a boundary iPEPS until convergence is reached. A single iPEPO layer absorption is performed by splitting it into a product of two-body gates which are contracted with the boundary iPEPS, followed by a  truncation similar to the one used in the SU imaginary time evolution algorithm. The method is applied twice in opposite directions to obtain an upper and a lower boundary iPEPS, representing the upper and lower half of the 3D network, respectively. The entire 3D network can then be effectively represented by a quasi-2D network made of the two boundary iPEPSs with a bulk iPEPO layer in between. The remaining 3-layer network is then contracted using CTMRG. Each of these stages of this algorithm, which we call the SU+CTMRG contraction, will be discussed in  detail in the following.

\subsection{SU approach for the  boundary iPEPS}
\begin{figure}
    \includegraphics[width=\linewidth]{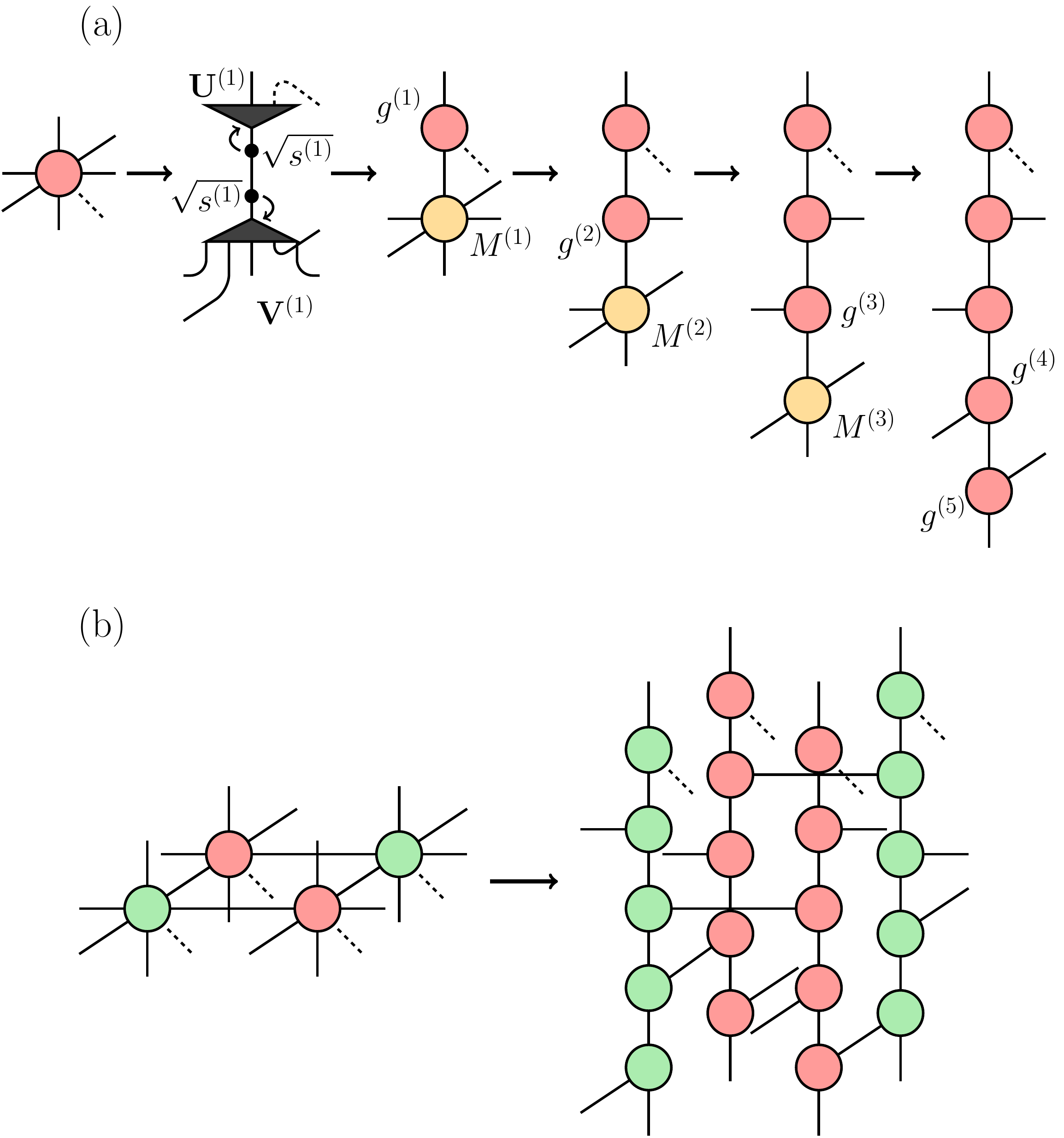}
    \caption{In (a) we show the decomposition of a bulk tensor in the 3D iPEPS to rank-3 tensors by sequentially splitting off legs using an SVD. After each SVD the square root of the singular value matrix is absorbed on each side (see main text for more details). In (b) the network of rank-3 tensors that is obtained after the decomposition is shown for a bipartite cubic lattice.}
    \label{fig:decompose_comb}
\end{figure}
We will start by explaining how an absorption of a single iPEPO layer onto the boundary iPEPS is performed. First, the bulk tensors are decomposed in such a way that the transformed network solely consists of rank-3 tensors. This is done by performing a sequence of SVDs in the following way (see also Fig.~\ref{fig:decompose_comb}(a))
\begin{align}
	b^{lrbfud}_p &= \mathbf{b}_{up,lrbfd} = \sum_{a_1} \mathbf{U}^{(1)}_{up,a_1} s^{(1)}_{a_1} \mathbf{V}^{* (1)}_{lrbfd,a_1} \nonumber\\
	&= \sum_{a_1} g^{(1)}_{ua_1p} \mathbf{M}^{(1)}_{ra_1,lbfd} \nonumber\\
	&= \sum_{a_1a_2} g^{(1)}_{ua_1p} \mathbf{U}^{(2)}_{ra_1,a_2} s^{(2)}_{a_2} \mathbf{V}^{*(2)}_{lbfd,a_2} \nonumber\\
	&= \cdots = \nonumber\\
	&= \sum_{a_1a_2a_3a_4a_5} g^{(1)}_{ua_1p} g^{(2)}_{ra_1a_2} g^{(3)}_{la_2a_3} g^{(4)}_{fa_3a_4} g^{(5)}_{ba_4d}
\end{align}
where $b^{lrbfud}_p$ is a tensor of the iPEPS network with the square root of the singular value matrices on the virtual bonds absorbed into it. At each decomposition step the square root of the singular value matrix obtained from the SVD is absorbed on each side, so we have $\mathbf{g}^{(j)} = \mathbf{U}^{(j)} \sqrt{s^{(j)}}$ and $\mathbf{M}^{(j)} = \sqrt{s^{(j)}} \mathbf{V}^{(j) \dagger}$. Note that the bond dimension in the middle of each string of rank-3 tensors can become large. In principle a truncation can be done on the singular value spectrum to reduce the computational cost. In practice, however, it turns out that the CTMRG procedure, that is discussed in the next section, typically gives the dominant contribution, therefore we do not perform a truncation here. 

\begin{figure}
    \includegraphics[width=1\linewidth]{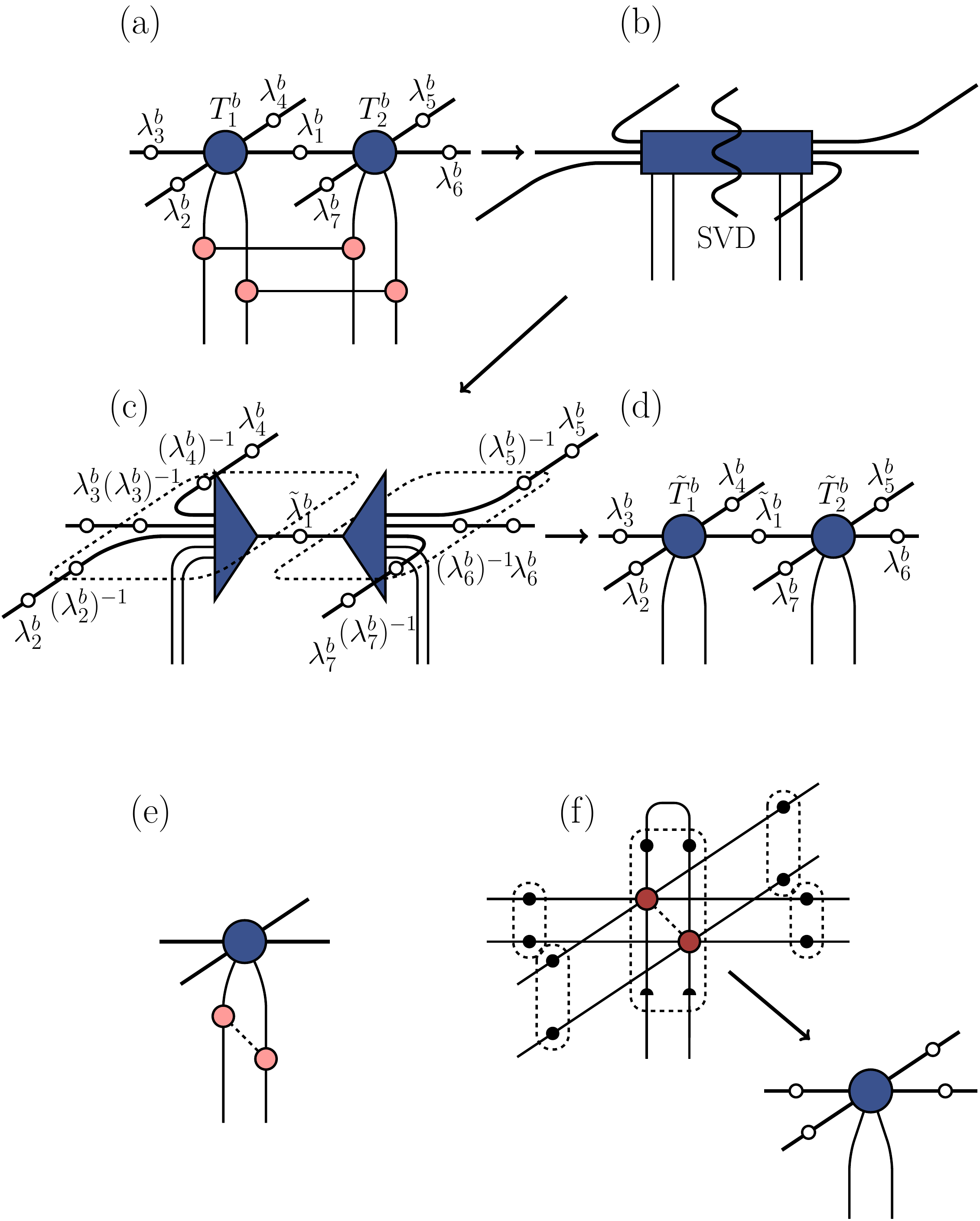}
    \caption{In (a-d) one step of the SU contraction is displayed. The open circles indicate the singular value matrices $\lambda_i^b$ that are introduced on the boundary iPEPS. In (a) tensors from the bra- and ket-layer are applied onto the boundary iPEPS tensors as are the singular value matrices from the neighboring bonds. In (b) an SVD is performed, after which the identities $\lambda_i^b (\lambda_i^b)^{-1}$ are inserted on each external bond in (c). In (d) the $(\lambda_i^b)^{-1}$ are contracted to obtain the new tensors. In (e) an absorption of the tensors carrying the physical bonds is shown which does not require a truncation. (f) shows the initialization of the upper boundary iPEPS tensors which are obtained from a contraction of the bulk iPEPS tensors. On the top, two singular value matrices of the bulk tensors, which are indicated by filled black circles, are used as an effective environment in a similar spirit as in the cluster contraction. The half circles indicate that a square root of the singular value matrix is taken. The initial singular value matrices of the boundary iPEPS are obtained by combining the bra- and ket-layer singular value matrices from the bulk tensors in the plane.}
    \label{fig:su_ctm_step_comb}
\end{figure}

Figure~\ref{fig:decompose_comb}(b) shows an example of the network that is obtained from this decomposition for a bipartite lattice. The tensors on the two sublattices are split in different orders, such that the connecting horizontal legs between neighboring pairs of tensors are properly aligned. By thinking of the pairs of rank-3 tensors as two-body gates, we can absorb them onto the boundary iPEPS and perform a truncation in a similar way as in the SU imaginary time evolution algorithm. A single step of this procedure is shown in detail in Fig.~\ref{fig:su_ctm_step_comb}(a-d). Both a gate from the bra- and the ket-layer are contracted at the same time. Figure~\ref{fig:su_ctm_step_comb}(e) shows the absorption of the tensors carrying the physical bond, which does not involve a truncation. The maximum bond dimension that is used for the boundary iPEPS is denoted by $\chi_b$. The iterative contraction of the iPEPO layers is continued until  convergence is reached in the computed expectation values, which for the benchmark models that will be discussed later is achieved in less than ten iterations. Alternatively, the convergence of the boundary iPEPS singular value matrices can also be used as a convergence criterion. In general, there is no reflection symmetry in the iPEPS. Therefore, a contraction from the opposite direction (using another boundary iPEPS) is also required with the mirrored procedure. The upper and lower boundary iPEPS are initialized by a contraction of the bulk tensors as shown in Fig.~\ref{fig:su_ctm_step_comb}(f). The dominant scaling of the boundary iPEPS contraction is $\mathcal{O}(\chi_b^5D^8 + \chi_b^4D^{10})$.

\subsection{3-layer CTMRG}

\begin{figure}
	\includegraphics[width=\linewidth]{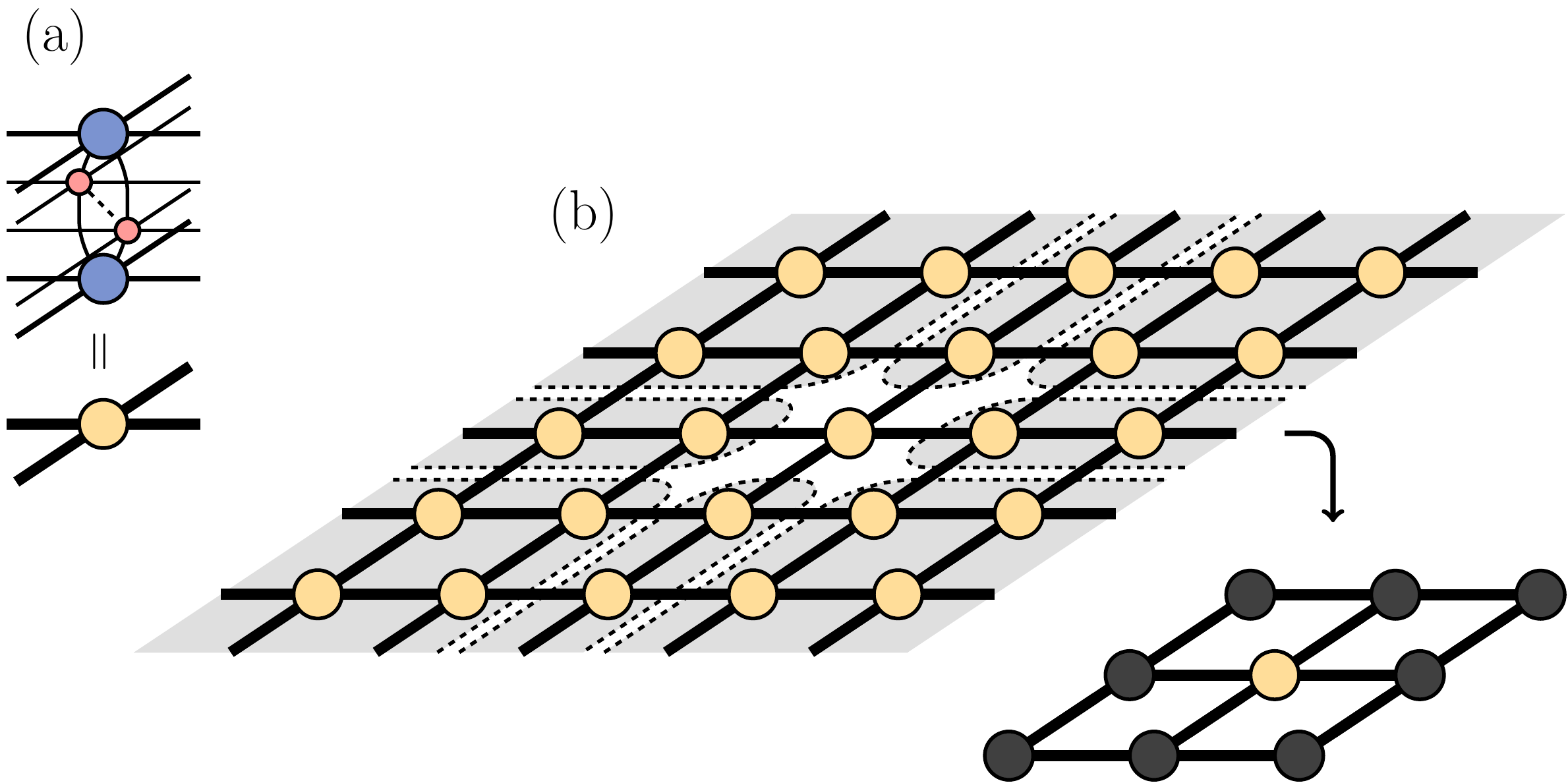}\vspace{5mm}
	\includegraphics[width=.913\linewidth]{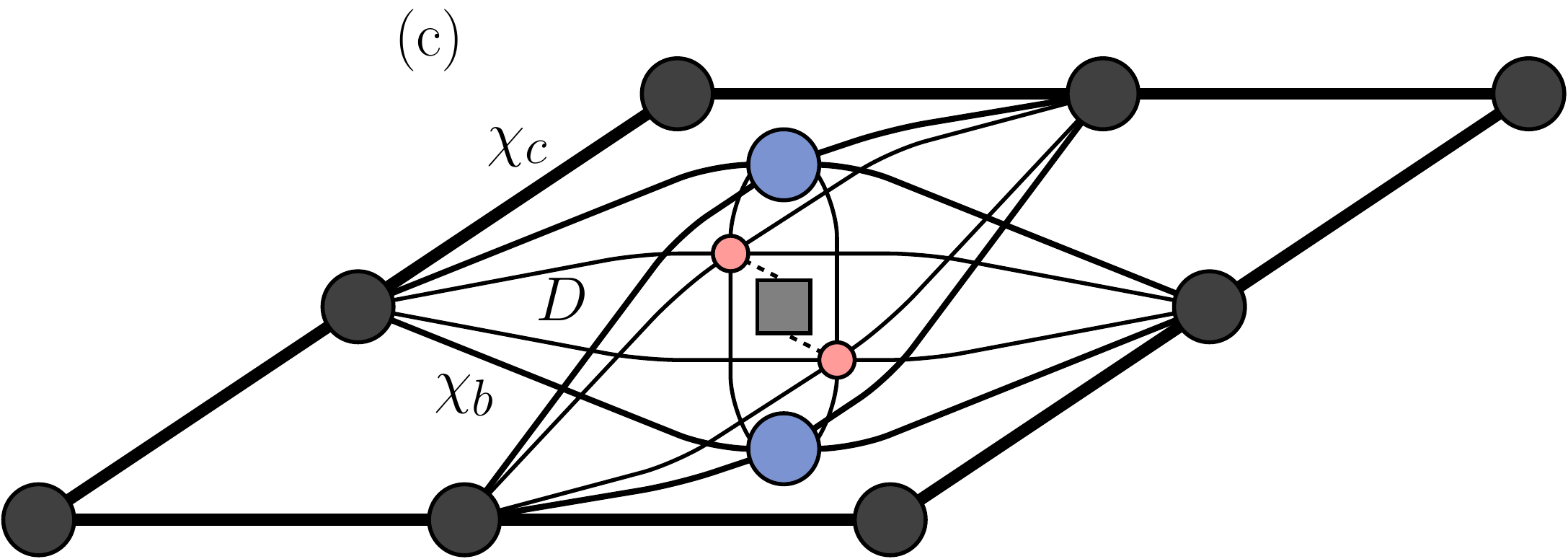}
	\caption{The 3D network is effectively represented by a three-layer network, made of the two converged boundary iPEPS with an iPEPO layer of bulk tensors sandwiched in between, shown in (a) for one site of the resulting quasi-2D network. This 2D network is contracted using the CTMRG yielding the corner and edge tensors surrounding the central site shown in (b), each representing part of the system as indicated by the shaded regions. In (c) the diagram corresponding to the expectation value of a one-site operator is displayed.}
	\label{fig:ctm_env}
\end{figure}

In the first part of the algorithm discussed in the previous section we explained how to obtain the converged upper and lower boundary iPEPS which represent the upper and lower half of the 3D network, respectively. The entire 3D network can then be represented by the double-layer network made of the two boundary iPEPS. For the computation of observables we additionally keep a single bulk iPEPO layer sandwiched between the two boundary iPEPS, resulting in the 3-layer network shown in Fig.~\ref{fig:ctm_env}(a-b). In the second part of the algorithm, this quasi-2D network is contracted using the CTMRG method which is a common approach for the contraction of 2D tensor networks~\cite{Nishino1996,Orus2009,Corboz2014}.

In the standard CTMRG algorithm~\cite{Nishino1996} the environment surrounding a central site of an infinite 2D network is approximated by four corner and four edge tensors, each representing an infinite quadrant and infinite row of tensors, respectively, as depicted in Fig.~\ref{fig:ctm_env}(b). These environment tensors are obtained through an iterative scheme in which rows and columns of tensors are absorbed in the environment tensors. In this work, the directional CTMRG~\cite{Orus2009}, with the renormalization procedure from Ref.~\cite{Corboz2014}, is used \footnote{We note that the QR decomposition in Ref.~\cite{Corboz2014} is not required for the computation of the projectors~\cite{Okubo}.}. The accuracy of the contraction is controlled by the bond dimension of the environment tensors, here denoted by $\chi_c$. The dominant scaling of this 3-layer CTMRG scheme is $\mathcal{O}(\chi_c^3 \chi_b^4 D^4 + \chi_c^2\chi_b^6D^6 + \chi_c^2\chi_b^4D^9)$, which makes this the computationally most expensive part of the SU+CTMRG contraction algorithm.

Once the environment tensors are converged, expectation values can be evaluated in the standard way, as shown in Fig.~\ref{fig:ctm_env}(c) for the example of a one-site operator. For two-site operators, only operators lying within the additional iPEPO layer can be directly computed. To evaluate two-site operators along the direction orthogonal to the iPEPO layer, another SU+CTMRG contraction is performed using a rotated iPEPS network.

\section{Results} \label{sec:results}

\subsection{Heisenberg model}

To benchmark the contraction methods, we first present results for the spin-$\frac{1}{2}$ anti-ferromagnetic Heisenberg model on the cubic lattice. This model is given by the Hamiltonian
\begin{equation}
	\hat{H} = J \sum_{\expval{i,j}} \hat{\mathbf{S}}_i \cdot \hat{\mathbf{S}}_j
\end{equation}
where $\hat{\mathbf{S}}_i$ are spin-$1/2$ operators and $J > 0$. The iPEPS tensors are obtained using the SU imaginary time evolution algorithm. We perform simulations for $D=2-4$ and, in order to reduce the computational cost, we use tensors with a $U(1)$ symmetry~\cite{Singh2011,Bauer2011}. 

\begin{figure}
	\includegraphics[width=\linewidth]{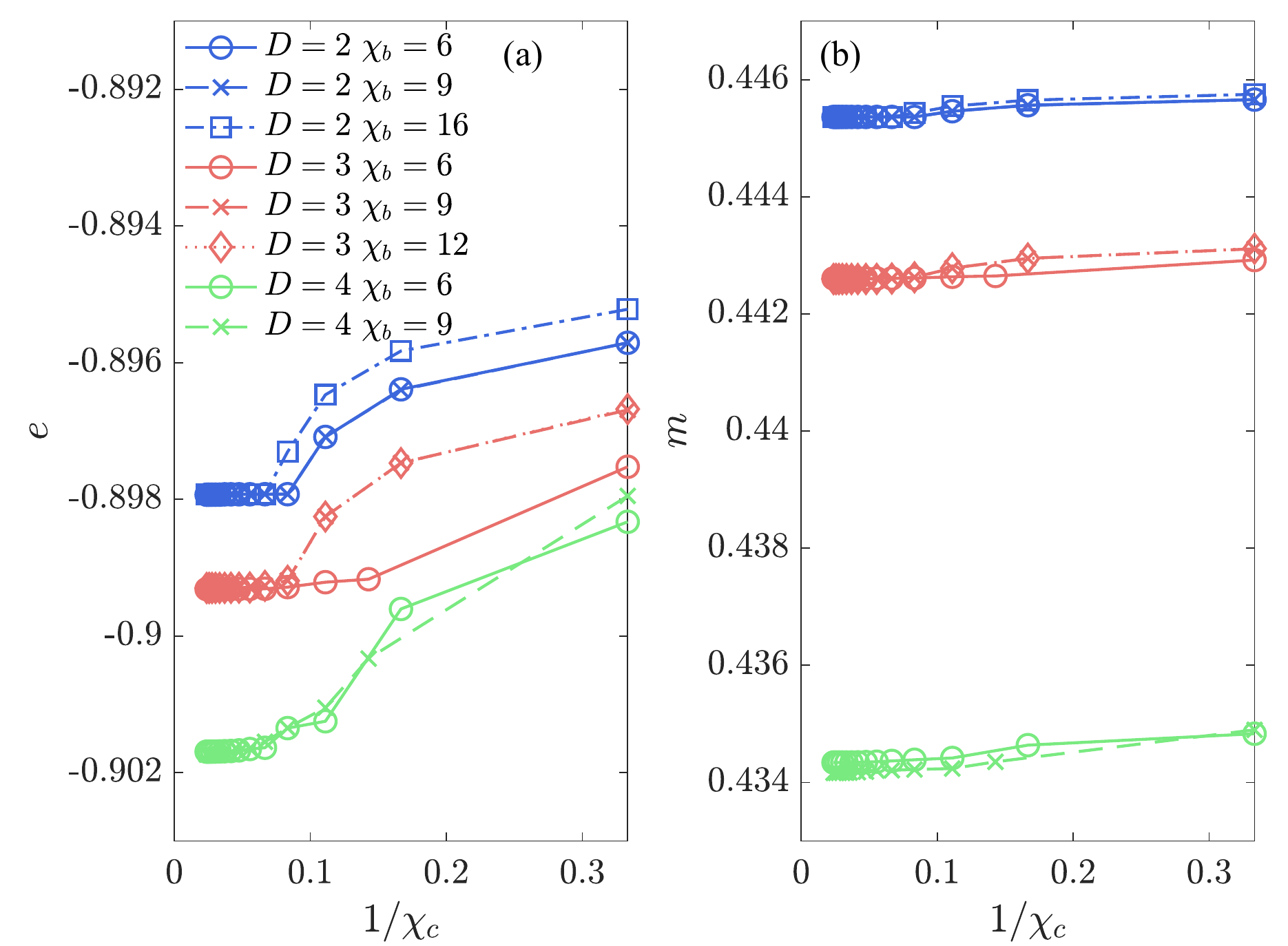}
	\caption{Convergence of (a) the energy and (b) the local magnetic moment $m$ as a function of inverse CTMRG boundary dimension $1/\chi_c$, for various $D$ and $\chi_b$.}
	\label{fig:heis_chic}
\end{figure}

We start by analyzing the convergence behavior of the SU+CTMRG contraction, first as a function of the CTMRG environment dimension $\chi_c$, for fixed values of $D$ and $\chi_b$. Figure $\ref{fig:heis_chic}$ presents results for the energy per site and the average local magnetic moment $m=\frac{1}{N}\sum_{i=1}^N \abs{\expval{\hat{\mathbf{S}}_i}}$, where $i$ goes over all the non-equivalent sites in the ansatz (i.e. $N=2$ in our ansatz with two independent tensors). For both observables convergence is achieved at moderate values of $\chi_c$, although higher values are required for larger $D$ and $\chi_b$, as is expected. In the following, the value of $\chi_c$ is fixed to a sufficiently large value, such that errors from the CTMRG contraction are negligible.

\begin{figure}
	\includegraphics[width=\linewidth]{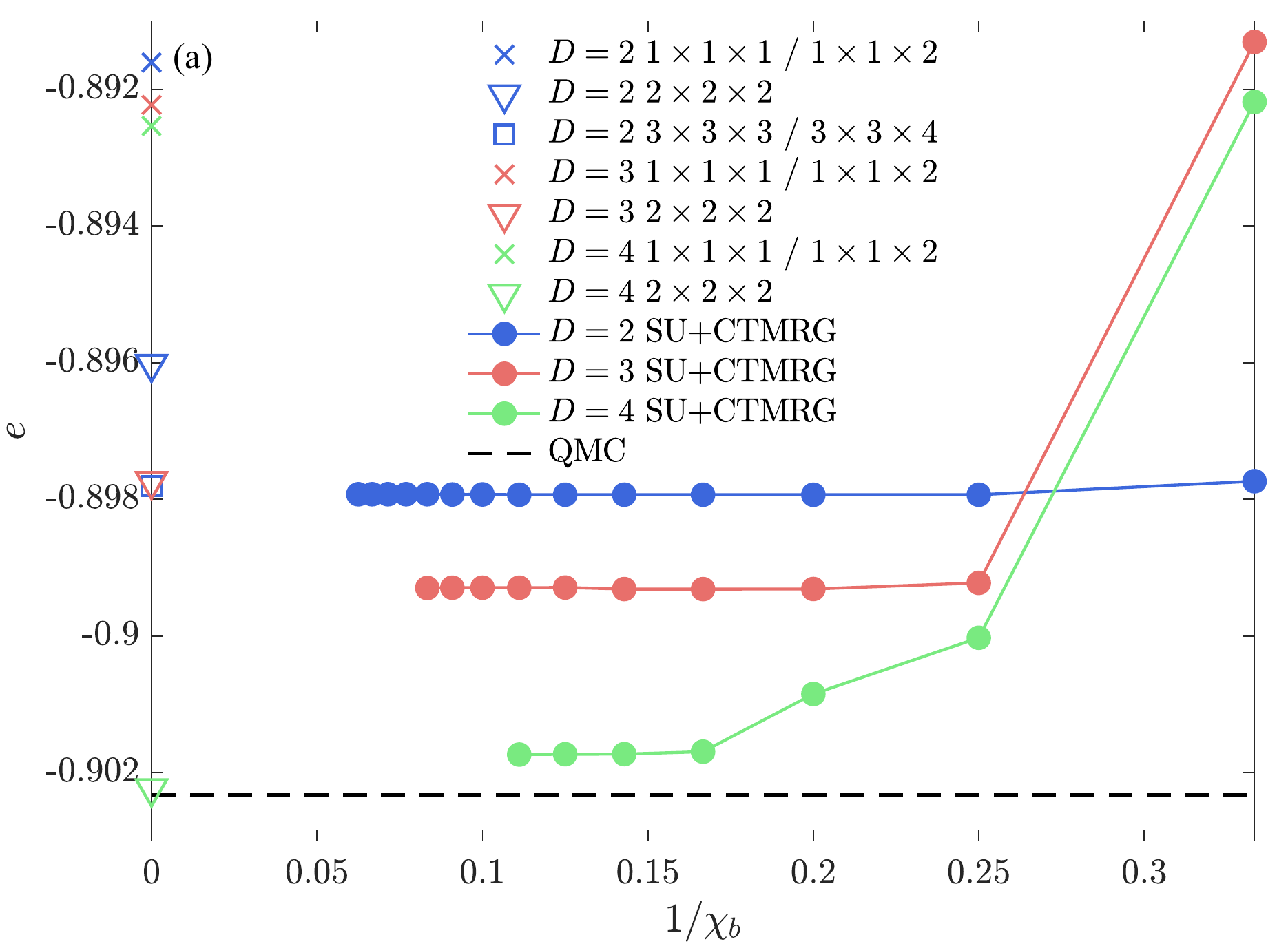}
	\includegraphics[width=\linewidth]{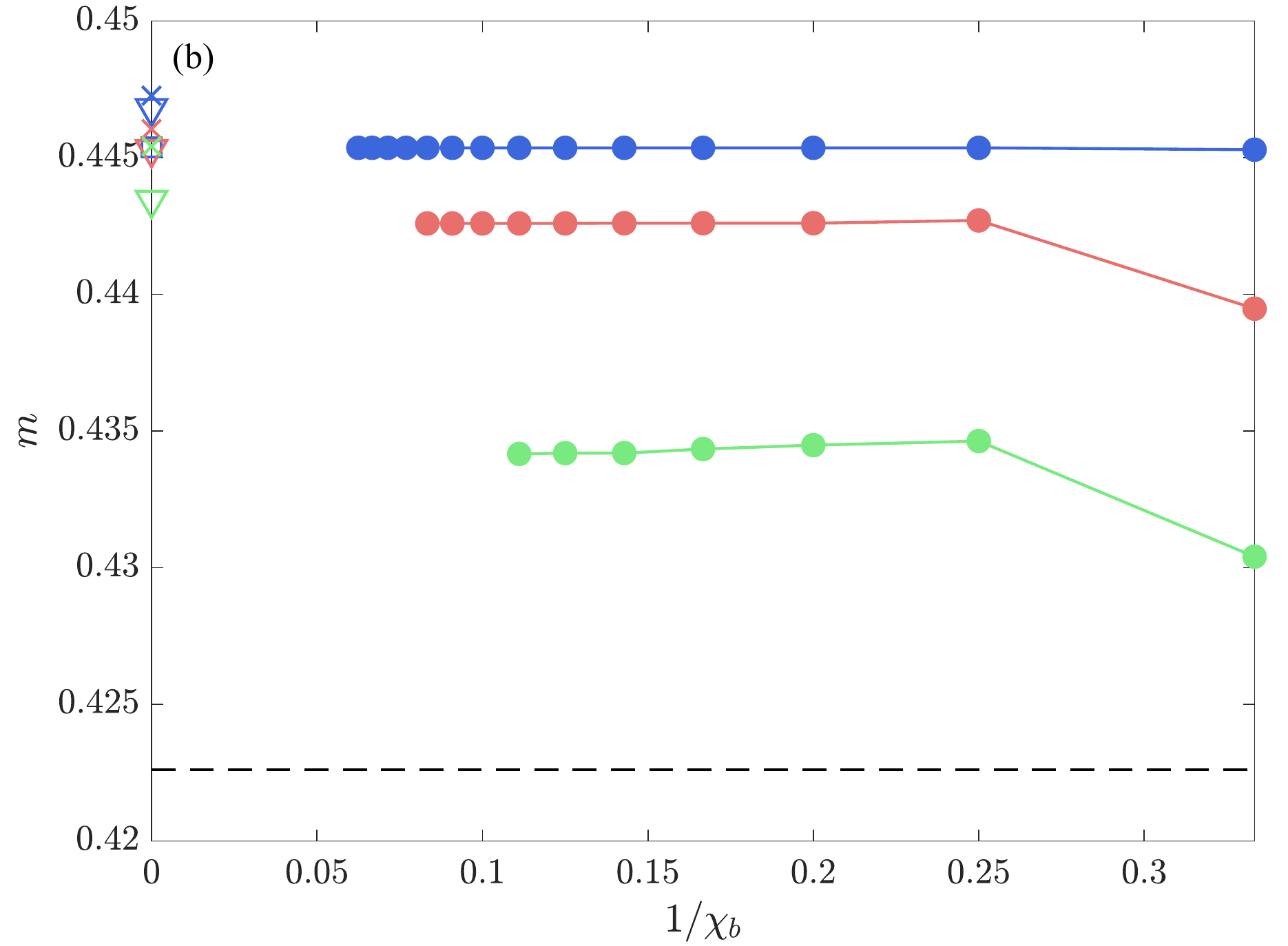}
	\includegraphics[width=\linewidth]{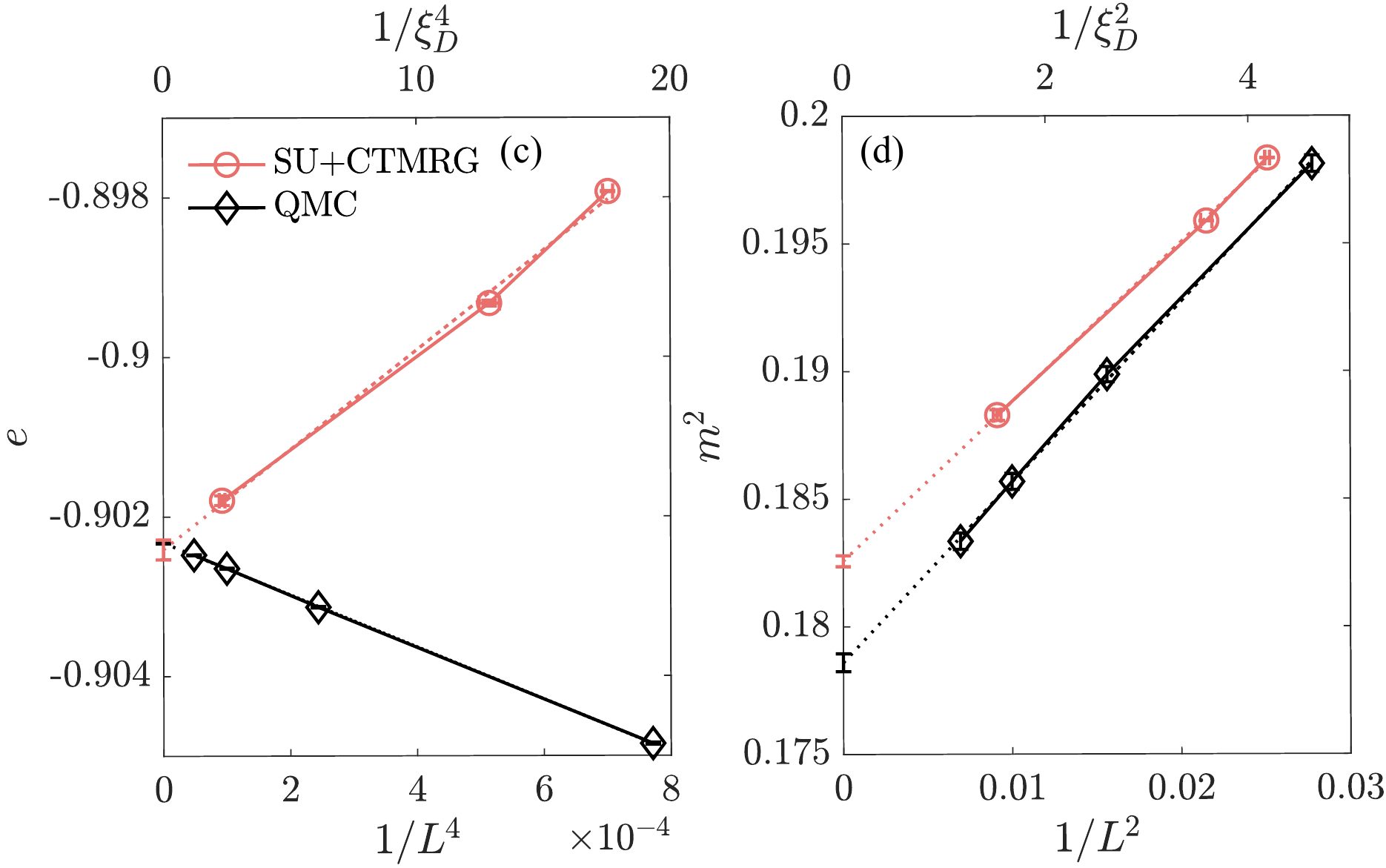}
	\caption{The results for (a) the energy and (b) the local magnetic moment $m$ obtained with the SU+CTMRG method as a function of the inverse boundary iPEPS dimension $1/\chi_b$ and the cluster contractions which are plotted on the vertical axis. The QMC result extrapolated to the thermodynamic limit is shown by the dashed line. In (c) and (d) a linear extrapolation based on the effective correlation length $\xi_D$ is presented for the energy as a function of $1/\xi_D^4$ and for $m^2$ as a function of $1/\xi_D^2$, respectively. For comparison, finite size scaling results from QMC using system sizes $L=6,8,10$ and $12$ are shown.}
	\label{fig:heis_chib}
\end{figure}

We next study the convergence of the SU+CTMRG as a function of the boundary iPEPS bond dimension, $\chi_b$, first focusing on the energy shown in Fig.~\ref{fig:heis_chib}(a). The SU+CTMRG contractions show a rapid convergence as a function of $\chi_b$. The $\chi_b$ required for convergence increases with $D$, still, even for the largest $D=4$ already a modest $\chi_b\sim 7$ results in a very small contraction error. In the same figure we also present data obtained from the cluster contractions. The energy obtained with the smallest $1\times1\times2$ cluster shows a large deviation from the SU+CTMRG results already for $D=2$, with only little improvement for higher $D$. The $2\times2\times2$ cluster gives a significant improvement over the $1\times1\times2$ cluster. The $3\times3\times4$ cluster improves this result further and shows a remarkable agreement with the SU+CTMRG result for $D=2$. As mentioned in Sec.~\ref{sec:cluster_contractions}, it was not possible to use this last cluster for higher values of $D$ due to the high computational cost. 

For the local magnetic moment $m$ shown in Fig.~\ref{fig:heis_chib}(b) similar observations can be made. The most important difference is that the $2\times2\times2$ cluster contraction provides a less significant improvement over the smallest $1\times1\times1$ cluster.

To obtain an estimate of the energy and $m$ in the exact infinite $D$ limit we attempt an extrapolation based on the effective correlation length $\xi_D$, an approach which has been used also for iPEPS in 2D~\cite{Rader2018,Corboz2018}. The main idea is that $\xi_D$ plays a similar role as a finite system size~\cite{Nishino1996(2),tagliacozzo08,pollmann2009,pirvu12} such that it can be used to perform a finite size scaling analysis.
The correlation length can be computed from the two leading eigenvalues of the transfer matrix represented by the edge tensors in CTMRG~\cite{Nishino1996(2)}, and $\xi_D$ corresponds to the value for a given $D$ in the $\chi_b,\chi_c \rightarrow \infty$ limit. We make use of the finite-size scaling ansatz for the energy and $m^2$ derived in Ref.~\cite{Hasenfratz1993}, where the leading finite-size corrections scale as $1/L^4$ and $1/L^2$, respectively. 

In Figs.~\ref{fig:heis_chib}(c) and (d) we present the iPEPS results for the energy and $m^2$ in comparison with QMC data, obtained with the loop algorithm from the ALPS library~\cite{Albuquerque2007,Bauer2011(2)} for system sizes up to $L=12$ and at sufficiently low temperatures ($T=0.005 J$) such that finite temperature effects are negligible compared to the error bars.
For the energy we find a good agreement between the extrapolated iPEPS, $-0.9024(1)$, and QMC result, $-0.902325(11)$. The estimate for $m^2$ obtained with iPEPS, $0.1826(2)$, is slightly higher than the QMC value, $0.1786(4)$, which is most likely due to the local SU optimization scheme used here (which typically tends to overestimate the order parameter). Still, the relative error is only $\approx 2\%$, and we expect that the accuracy can be further improved by using more accurate optimization schemes.

\begin{figure}
	\includegraphics[width=\linewidth]{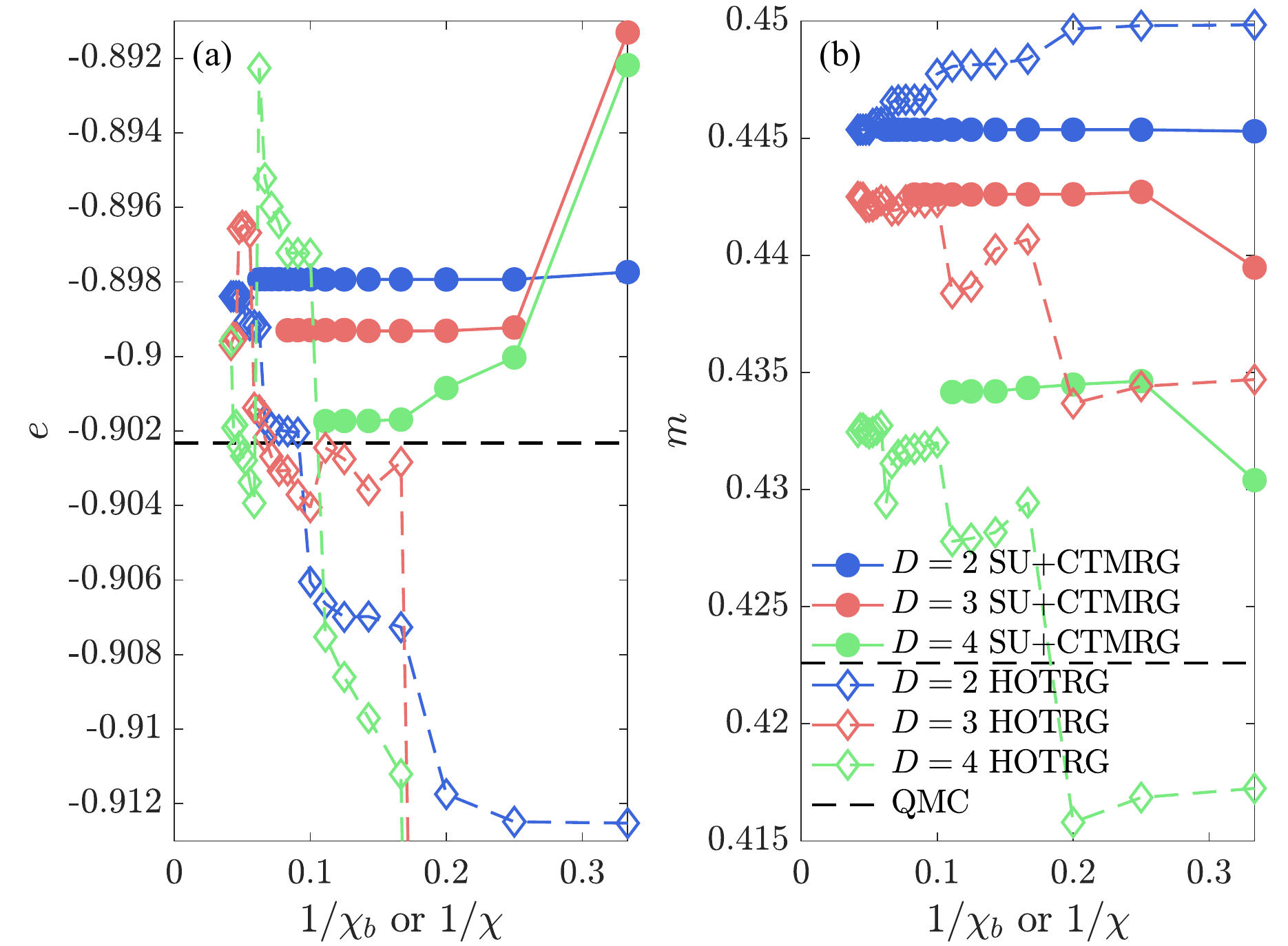}
	\caption{The HOTRG  results for (a) the energy and (b) the local magnetic moment $m$ as a function of the inverse bond dimension $1/\chi$ compared to the SU+CTMRG results as a function of $1/\chi_b$.}
	\label{fig:heis_chib_hotrg}
\end{figure}

Finally, in Fig.~\ref{fig:heis_chib_hotrg} we compare our SU+CTMRG results to the ones obtained with HOTRG which was previously proposed as a 3D (and 2D) contraction method in Ref.~\cite{Xie2012}. The main idea of HOTRG is to iteratively coarse-grain the tensor network in all spatial directions where the accuracy is controlled by the bond dimension $\chi$ of the coarse-grained tensors. Here we use a modified approach adapted to the anisotropic case where the projectors are computed in a similar way as in Ref.~\cite{Iino2019}. We observe that the convergence of the HOTRG results is strongly irregular and exhibits several plateaus, in contrast to the SU+CTMRG results which exhibit a fast and regular convergence. One possible reason for this behavior is that the distribution of the singular values obtained in HOTRG decays only very slowly, much slower than the spectrum in the SU+CTMRG approach. While there seems to be a tendency that HOTRG approaches the SU+CTMRG results, it was  not possible to reach high enough $\chi$ to fully converge due to the high computational cost. 

\subsection{Bose-Hubbard model}

As a second benchmark case we consider the Bose-Hubbard model defined by the Hamiltonian
\begin{equation}
	\hat{H} = -t\sum_{\expval{i,j}} \hat{b}_i^{\dagger}\hat{b}_j + \frac{U}{2} \sum_i \hat{n}_i(\hat{n}_i-1) - \mu\sum_i\hat{n}_i
\end{equation}
with $t$ the hopping amplitude, $U$ the on-site interaction, $\mu$ the chemical potential, $\hat{b}_i^{\dagger}$ ($\hat{b}_i$) the bosonic creation (annihilation) operator and $\hat{n}_i = \hat{b}_i^{\dagger}\hat{b}_i$ the number operator. At zero temperature the model exhibits Mott insulating phases with integer particle filling for $t \ll U$ and a superfluid phase (SF) for $t \gg U$~\cite{Fisher1989}.

To perform the iPEPS simulations with finite local Hilbert spaces we introduce a cutoff on the maximum occupation number on each site. The size of this cutoff is chosen such that the induced error is negligible. For the simulations of the $n=1$ and $n=2$ Mott lobes, a cutoff of $n_{\textup{max}}=3$ and $n_{\textup{max}}=4$ are used, respectively. To obtain the data in one of the phases, simulations are started deep in this phase and the converged iPEPS at one datapoint is used as the initial state of the SU optimization at the next datapoint. For a given value of $D$, the phase transition point can be determined by locating the intersection  of the iPEPS energies of the two phases. In contrast to the Heisenberg case in the previous section, the $U(1)$ symmetry cannot be exploited here because it is broken in the superfluid phase. For this reason the maximal bond dimension we consider here is $D=3$. The SU+CTMRG contractions are performed using $\chi_b=8$ and $\chi_c=21$, which are sufficiently large to make the remaining finite $\chi_b$ and $\chi_c$ errors much smaller than the symbol sizes.

\begin{figure}
	\includegraphics[width=\linewidth]{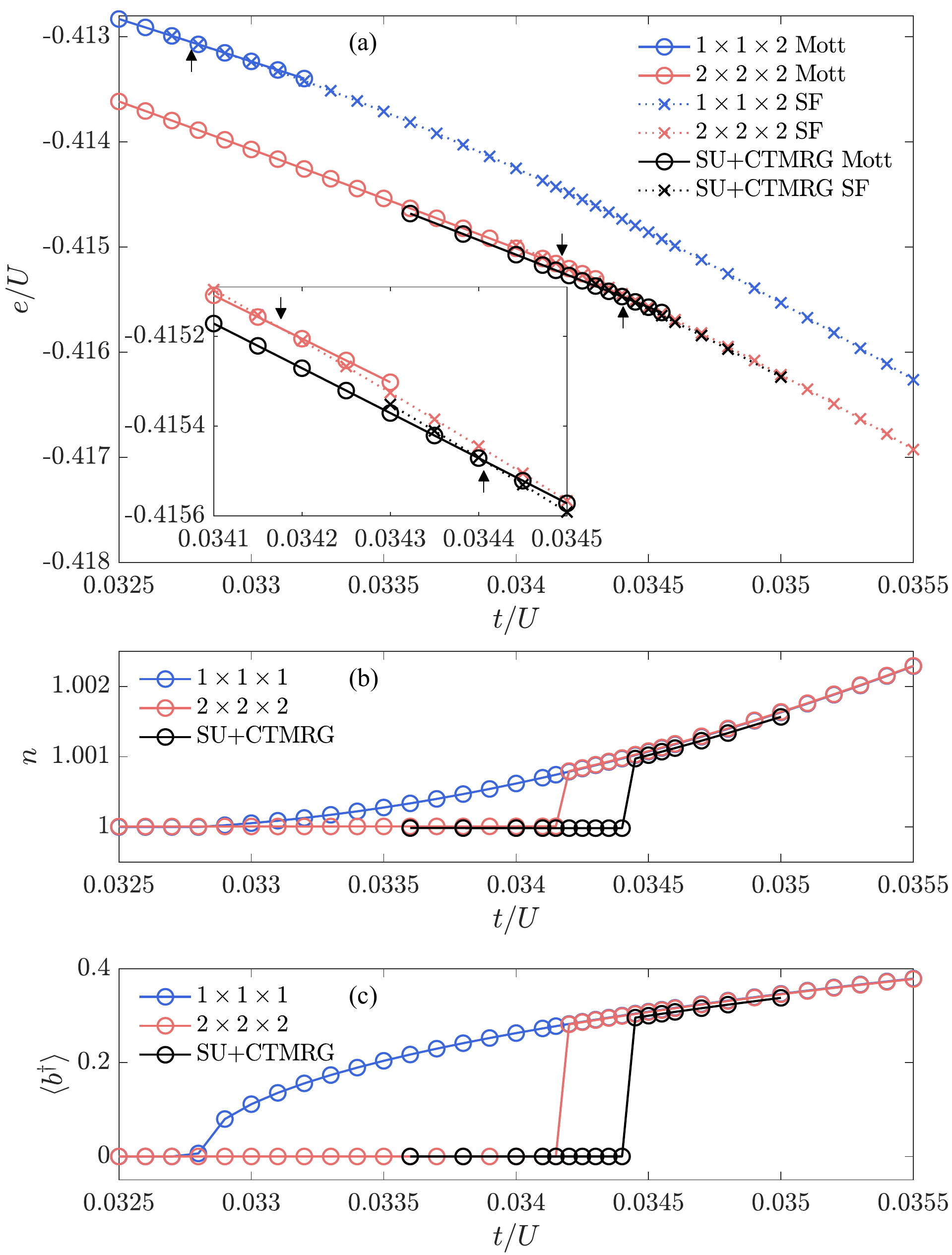}
	\caption{A cut through the phase diagram at the tip of the first Mott lobe for fixed $\mu/U=0.4$ as a function of $t/U$ obtained with $D=3$. In (a) the energy of both phases is shown obtained using the cluster and SU+CTMRG contraction. The inset displays a zoom of the same results close to where the contractions indicate the phase transition to take place, marked by arrows. In (b) and (c) the particle number and order parameter $\expval{b^{\dagger}}$ of the lowest energy state are shown, respectively.}
	\label{fig:bh_cut_mu4_U10}
\end{figure}

\begin{figure}
	\includegraphics[width=\linewidth]{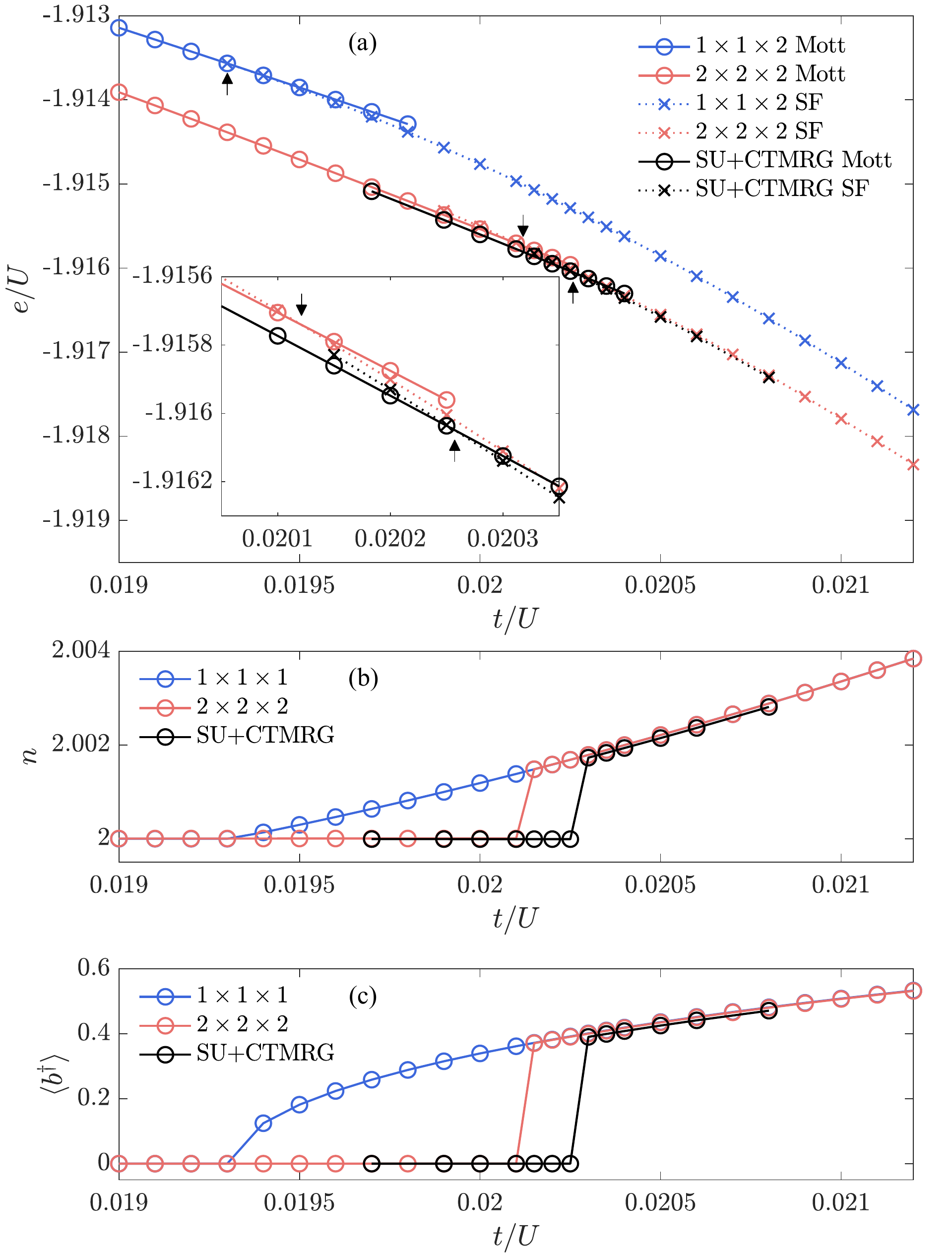}
	\caption{Same as in Fig.~\ref{fig:bh_cut_mu4_U10} for a cut through the phase diagram at the tip of the second Mott lobe with fixed $\mu/U=1.45$.}
	\label{fig:bh_cut_mu14,5_U10}
\end{figure}

We first consider two selected cuts in the phase diagram at the tip of the first and second Mott lobe. Figure~\ref{fig:bh_cut_mu4_U10} shows a comparison between results obtained from the SU+CTMRG and cluster contractions for the energy, particle number and the order parameter $\expval{b^{\dagger}}$ close to the tip of the $n=1$ lobe for fixed $\mu/U=0.4$ as a function of $t/U$. For the energy, the $1\times1\times2$ cluster shows a large deviation compared to the other results. The $2\times2\times2$ cluster gives a significant improvement and there is only a slight shift in the location of the phase transition compared to the SU+CTMRG contraction (which is mostly due to the small angle at which the energies of the two phases intersect). For the particle number and the order parameter a much smaller improvement is seen when going from the $1\times1\times1$ cluster to the $2\times2\times2$ cluster   when compared to the SU+CTMRG result, as previously observed for $m$ in the Heisenberg model. Still, the absolute error on the scale shown in Figs.~\ref{fig:bh_cut_mu4_U10}(b-c) is small. Figure~\ref{fig:bh_cut_mu14,5_U10} shows results close to the tip of the second Mott lobe at fixed $\mu/U=1.45$, for which similar observations can be made.

We note that the jump in the order parameter $\expval{b^{\dagger}}$ at the phase transition in the $2\times2\times2$ cluster and the SU+CTMRG results seems to indicate that the transition is of first order, whereas the transition is known to be of second order. This is most likely an artifact of the SU optimization scheme used here which, since it is a local update, does not accurately reproduce the diverging correlation length across a second order transition. The accuracy of the order parameter around the transition could be improved using more accurate optimization schemes~\cite{Jordan2008,Phien2015,Wang2011,Liao2019}. Still, we can nevertheless obtain an accurate estimate of the critical point based on the intersection of the energies when compared to QMC as we show in the following.

\begin{figure}
	\includegraphics[width=\linewidth]{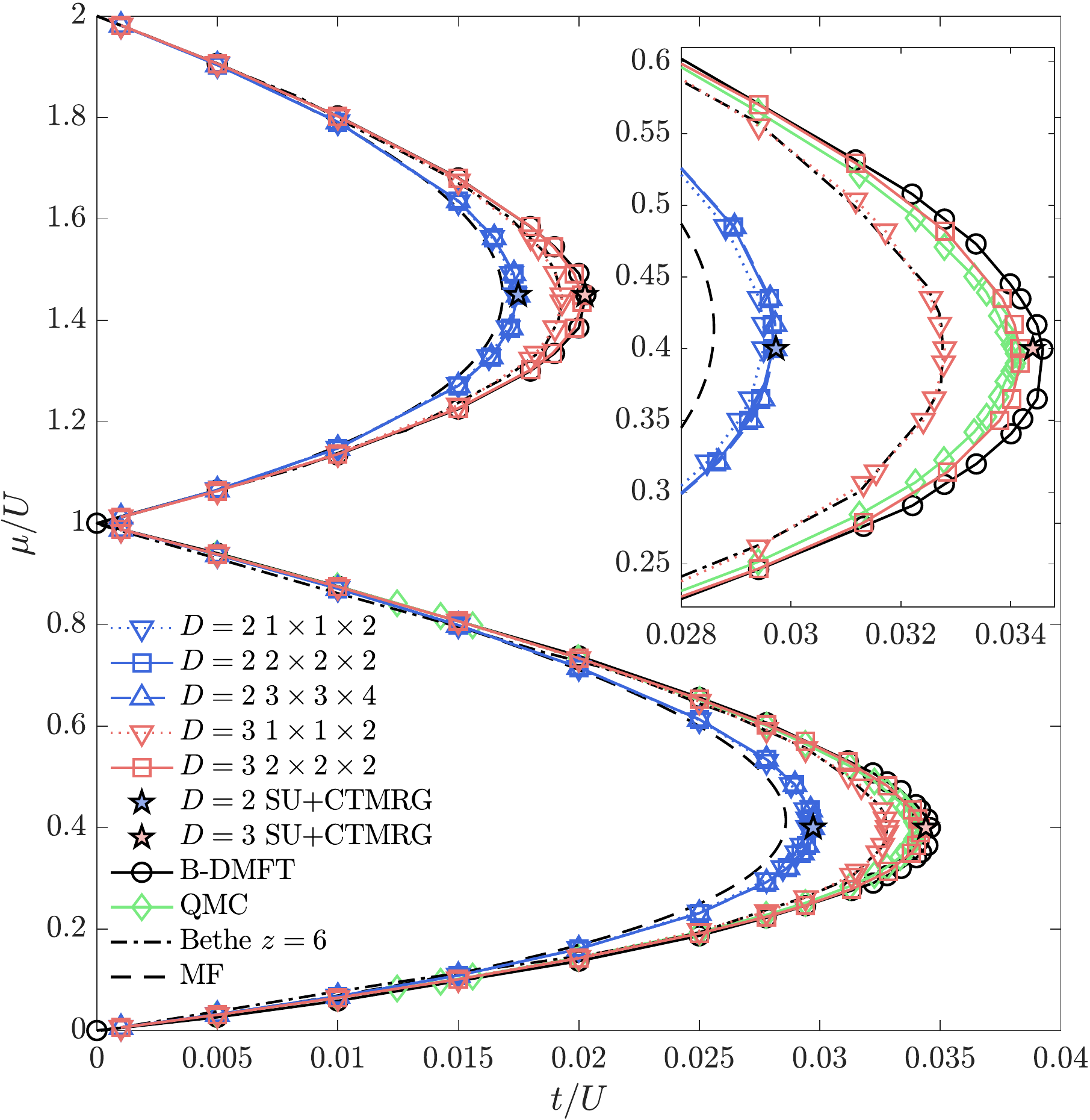}
	\caption{Ground state phase diagram of the first two Mott lobes of the Bose-Hubbard model obtained with the cluster and SU+CTMRG contraction. For comparison, results from B-DMFT~\cite{Anders2010}, QMC~\cite{Capogrosso-Sansone2007}, an exact solution for a Bethe lattice with coordination number $z=6$~\cite{Semerjian2009}, and static MF theory are shown.}
	\label{fig:phase_diagram}
\end{figure}

By performing simulations along additional cuts we can construct the ground state phase diagram for the first two Mott lobes shown in Fig.~\ref{fig:phase_diagram}. We have restricted these additional simulations to the cluster contractions, which are computationally substantially cheaper than the SU+CTMRG method, and, as we have found previously, the $2\times2\times2$ cluster already provides an estimate which is close to the SU+CTMRG result. For comparison, we added previous results from bosonic dynamical mean-field theory (B-DMFT)~\cite{Anders2010}, QMC~\cite{Capogrosso-Sansone2007}, an exact solution on the Bethe lattice for coordination number $z=6$~\cite{Semerjian2009}, and the mean-field result (MF). While the $D=2$ phase boundary only provides a slight improvement over the MF result, the data obtained for $D=3$ with the $2\times2\times2$ cluster and the SU+CTMRG contraction show a close agreement with the QMC results. This demonstrates that already a relatively small bond dimension is sufficient to obtain the phase diagram with a remarkable accuracy that is competitive or even better than B-DMFT. Another observation we can make is the agreement between the $D=3$ results obtained by the $1\times1\times2$ cluster and the $z=6$ Bethe lattice results. This is because the simulation based on the SU imaginary time evolution approach combined with the $1\times1\times2$ contraction is equivalent to an iPEPS simulation on the Bethe lattice.

\section{Summary and discussion} \label{sec:discussion}

In this paper we have presented two iPEPS contraction approaches to study 3D quantum many-body systems. The cluster contraction provides an approximation to the 3D contraction by only contracting a small cluster of tensors exactly while the rest of the network is taken into account approximately via the singular values on the boundary bonds of the cluster. The contraction error for the smallest $1\times1\times1$ and $1\times1\times2$ clusters, which have been used in previous studies~\cite{Picot2015,Picot2016,Jahromi2019,Jahromi2020}, can be quite large. A considerable improvement (at least for the energy) is obtained when using the larger $2\times2\times2$ cluster which is computationally still affordable. The SU+CTMRG method enables a full contraction of the 3D tensor network by iteratively absorbing iPEPO layers with a lower and upper boundary iPEPS, and by contracting the resulting quasi-2D tensor network using CTMRG. The accuracy can be systematically controlled by $\chi_b$ and $\chi_c$, the bond dimension of the boundary iPEPS and the CTMRG environment tensors, respectively. A fast convergence as function of $\chi_b$ and $\chi_c$ is observed, significantly outperforming a HOTRG contraction for the Heisenberg model. For the Bose-Hubbard model we found that already a relatively small $D=3$ yields a phase diagram which is in close agreement with QMC. We have shown that the combination of the two contraction approaches provides a practical way to compute a phase diagram at a reasonable computational cost.

There are various ways in which the accuracy of the methods can be further improved. For the optimization of the iPEPS tensors a higher accuracy can be obtained by using a full~\cite{Jordan2008}, or fast-full update~\cite{Phien2015}, which, for each truncation step in the imaginary time evolution, take the entire wave function into account, in contrast to the local SU update. However, the  computational cost of these approaches is also substantially higher than the SU scheme. Alternatively, a cluster update~\cite{Wang2011,Lubasch2014,Lubasch2014(2)}, which only takes a cluster of tensors into account to perform a truncation, may provide an optimal trade-off between accuracy and computational cost. Schemes based on a direct energy minimization, e.g. based on automatic differentiation~\cite{Liao2019}, may  be another interesting option. The SU+CTMRG contraction itself could in principle also be further improved by replacing the SU by a full (or cluster) update scheme, albeit also here at the expense of a higher computational cost. These topics will be explored in future work.

We expect these methods to provide a promising path towards simulating challenging 3D quantum systems, such as e.g. the pyrochlore Heisenberg model, layered systems, and ultra-cold atoms in optical lattices, especially for cases which are out of reach by QMC due to the negative sign problem. Finally, we note that these approaches can also be extended to the finite temperature case or to other types of lattices in a rather straightforward way, as was done in 2D.

\vspace{0.5cm}

\begin{acknowledgements}
We thank G. Vidal for useful discussions. This project has received funding from the European Research Council (ERC) under the European Union's Horizon 2020 research and innovation programme (grant agreement No 677061). This work is part of the D-ITP consortium, a program of the Netherlands Organization for Scientific Research (NWO) that is funded by the Dutch Ministry of Education, Culture and Science (OCW).
\end{acknowledgements}

\bibliography{references}

\end{document}